\documentclass[a4paper,fleqn,usenatbib]{mnras}

\usepackage{paralist,xcolor}

\usepackage[T1]{fontenc}
\usepackage{ae,aecompl}
\usepackage[normalem]{ulem} 

\usepackage{graphicx}	
\usepackage{amsmath}	
\usepackage{amssymb}	
\usepackage{pdflscape}
\usepackage{longtable}
\usepackage{booktabs}
\usepackage{rotating}
\usepackage{placeins}

\title[Continued  neutron star crust cooling in IGR\,J17480--2446]{Continued cooling of the accretion-heated  neutron star crust in the X-ray transient IGR\,J17480--2446 located in the globular cluster Terzan 5}

\author[L. S. Ootes et al.]{L. S. Ootes$^1$\thanks{E-mail (corresponding author): l.s.ootes@uva.nl},
S. Vats$^1$,
D. Page$^2$, R. Wijnands$^1$, A. S. Parikh$^1$,
N. Degenaar$^1$,  
\newauthor M. J. P. Wijngaarden$^3$, D. Altamirano$^4$, A. Bahramian$^5$, E. M. Cackett$^6$,  
\newauthor C. O. Heinke$^7$, J. Homan$^{8,9}$, J. M. Miller$^{10}$
\\
$^1$ Anton Pannekoek Institute for Astronomy, University of Amsterdam, Science Park 904, 1098 XH Amsterdam, The Netherlands\\
$^2$ Instituto de Astronom\'{i}a, Universidad Nacional Aut\'{o}noma de Me\'{x}ico, Mexico D.F. 04510, Mexico\\
$^3$ Mathematical Sciences, University of Southampton, SO17 1BJ, Southampton, United Kingdom\\
$^4$ Physics and Astronomy, University of Southampton, SO17 1BJ, Southampton, United Kingdom\\
$^5$ Department of Physics and Astronomy, Michigan State University, East Lansing, MI, USA\\
$^6$ Department of Physics and Astronomy, Wayne State University, 666 W. Hancock St. Detroit, MI 48201, USA\\
$^7$ Department of Physics, University of Alberta, CCIS 4-183, Edmonton, AB T6G 2E1, Canada\\
$^8$ Eureka Scientific Inc., 2452 Delmer Street, Oakland, CA 94602, USA\\
$^9$ SRON, Netherlands Institute for Space Research, Sorbonnelaan 2, 3584 CA Utrecht, The Netherlands \\
$^{10}$ Department of Astronomy, University of Michigan, 1085 South University Ave., Ann Arbor, MI 48109, USA}

\date{Accepted XXX. Received YYY; in original form ZZZ}

\pubyear{2018}

\begin{document}
\label{firstpage}
\pagerange{\pageref{firstpage}--\pageref{lastpage}}
\maketitle

\begin{abstract} 
We present a new {\it Chandra} observation (performed in July 2016) of the  neutron star X-ray transient IGR\,J17480--2446, located in the globular cluster Terzan 5. We study the continued cooling of the neutron star crust in this system that was heated during the 2010 outburst of the source. This new observation was performed two years after the last observation of IGR\,J17480--2446, hence, significantly extending the cooling baseline. We reanalysed all available {\it Chandra} observations of the source (but excluding observations during which one of the known transients in Terzan 5 was in outburst) and fitted the obtained cooling curve with our cooling code {\sc NSCool}, which allows for much improved modelling than what was previously performed for the source. The data and our fit models indicate that the crust was still cooling $\sim$5.5~years after the outburst ended. The neutron star crust has likely not reached crust-core thermal equilibrium yet, and further cooling is predicted (which can be confirmed with additional {\it Chandra} observations in $>5$~years). Intriguingly, we find indications that the thermal conductivity might be relatively low in part of the crust compared to what has been inferred for other crust-cooling sources and tentatively suggest that this layer might be located around the neutron drip. The reason for this difference is unclear, but might be related to the fact that IGR\,J17480--2446 harbours a relatively slowly rotating neutron star (with a spin of 11~Hz) that has a relatively strong inferred surface magnetic field ($10^{9-10}$~Gauss) compared to what is known or typically assumed for other cooling sources.
\end{abstract}




\section{Introduction}
\label{section:p4:intro}

Low-mass X-ray binaries (LMXBs) harbouring neutron star primaries are excellent laboratories for studying the thermal evolution of accretion-heated  neutron star crusts and, through such studies, they allow us to investigate the physical processes at work in the dense matter present in such crusts. As matter gets accreted from the (sub--)solar secondary (via Roche-lobe overflow) onto the surface of the neutron star, the newly accreted matter compresses the underlying material in the stellar crust. This leads to a series of exothermic reactions like electron captures, neutron emission and, most energetically, density driven nuclear fusion reactions \citep[e.g.][]{haensel1990dch,haensel2008,steiner2012} that cause heating of the  neutron star crust and eventually the  neutron star core \cite[e.g.][]{bbr1998,colpi2001}. Most of the energy generated by these reactions is released deep in the crust (at a density of ${\sim 10^{12-13}\text{ g cm}^{-3}}$) and this process is therefore commonly referred to as the ``deep crustal heating" mechanism.

A few dozen neutron star LMXBs are persistently accreting sources so that the neutron star is typically outshone by the  X-rays produced by the accretion process. However, many neutron star LMXBs systems are transients that only occasionally accrete matter during so-called outbursts \citep[for a review on neutron star X-ray transients see][]{campana1998}. The duration of such outbursts can vary widely and can range anywhere from several days \citep[e.g.][]{wijnands2009,heinke2010,matasanchez2017} to decades (e.g. \citealt{remillard1999,wijnands2001}; for reviews on the outburst behaviour of X-ray transients we refer to \citealt{chen1997} and \citealt{yan2015}; for a review about the physical processes behind such outbursts see \citealt{lasota2001}). During an outburst, the X-ray luminosities of transient neutron star LMXBs reach $L_\text{X}\sim$10$^{35}$--10$^{38}$~erg\,s$^{-1}$. Such accretion outbursts are followed by longer periods of quiescence. During this phase, they can still be detectable at X-ray luminosities of $L_\text{X}\sim$10$^{31}$--10$^{34}$~erg\,s$^{-1}$. This low level of the quiescent X-ray luminosity indicates that the accretion onto the neutron star is only active at a very low rate or has almost completely halted. In the latter case, if the neutron star is sufficiently hot, we are able to study the emission directly from the neutron star surface, from which the thermal state of the neutron star (both its crust and its core) can be inferred \citep[for a review focusing on the observational results on this topic see][]{wijnands2017}.

During outburst, the energy released from the aforementioned reactions heats up the crust. If a sufficient amount of energy is released, the crust can be heated out of thermal equilibrium with the core. After the end of the outburst, the crust will cool until it is in equilibrium with the core again \citep[e.g.][]{bbr1998,ushomirsky2001,rutledge2002ks}. This cooling has been observed for about ten\footnote{For a few sources it needs to be confirmed that they are indeed crust-cooling sources as the observational results for them are not yet conclusive \citep[see][]{wijnands2017}.} systems \citep[][]{wijnands2017} and such studies have led to new insights into the physical processes at work in the crust. The two highest impact conclusions from these studies are that the crust has a high thermal conductivity (\citealt{wijnands2002,cackett2006,shternin2007,brown2009}; while, before these studies, typically a much lower conductivity was expected; e.g. \citealt{schatz2001}) and that, besides the deep crustal heating, there must be an additional source of heating present at much lower depths in the crust ($\rho\sim10^9\text{ g cm}^{-3}$) in several of the studied sources (e.g. \citealt{brown2009,degenaar2011ter5-418,homan2014,waterhouse2016})\footnote{Several studies of thermonuclear explosions (so-called type-I X-ray bursts) on the surface of the neutron star also, independently, inferred the existence of this shallow heat source \citep[e.g.][]{cumming2006,zand2012}.}. Currently the origin of this shallow heat source is not understood \citep[this is typically referred to as the ``shallow  heating problem"; for an in-depth discussion see][]{deibel2015}. 

\begin{figure}
\label{fig:p4:fov}
\includegraphics[clip=,width=1.0\columnwidth]{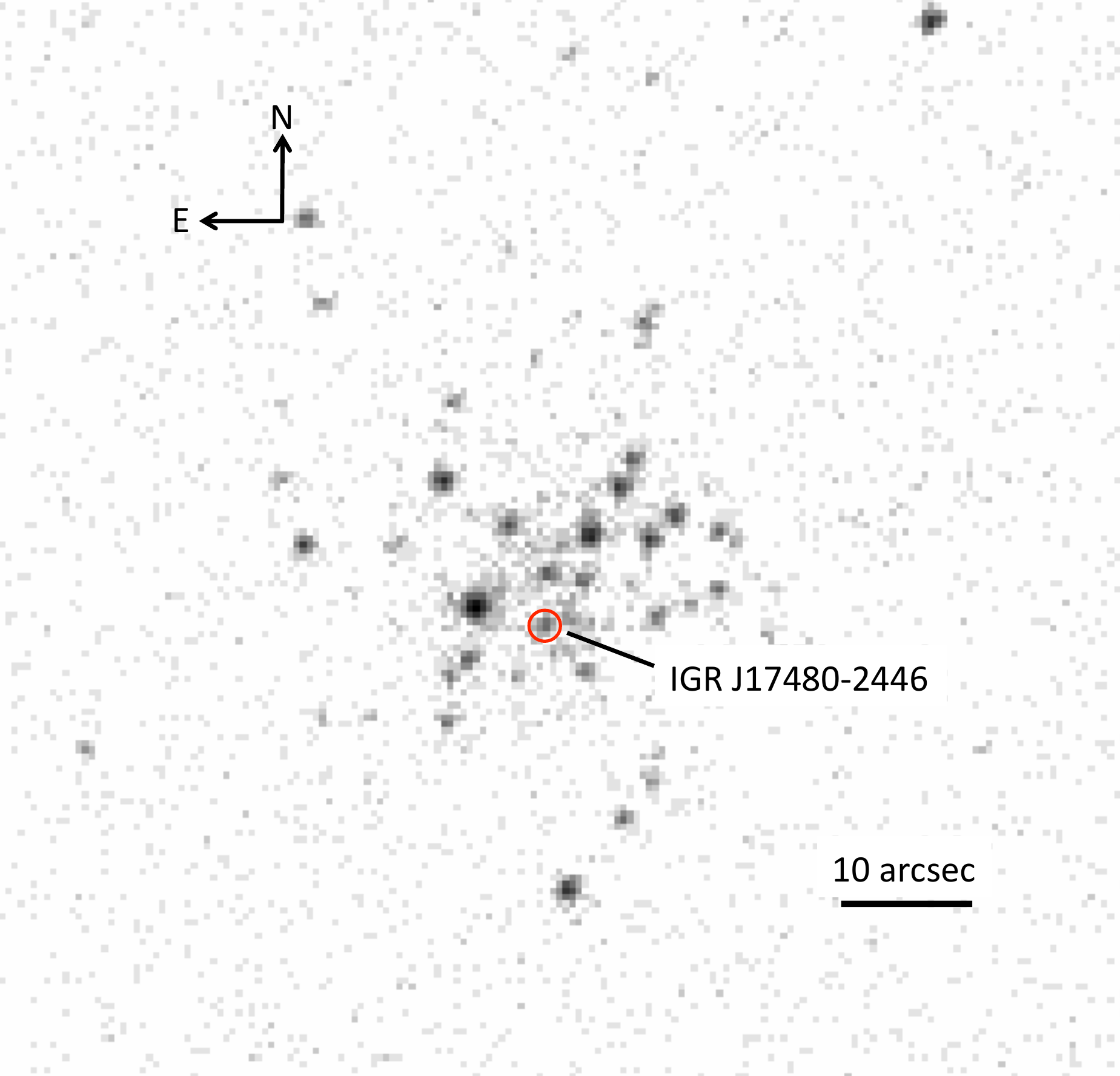}
\caption{The {\em Chandra} 0.5--10 keV X-ray image of the globular cluster Terzan\,5 obtained in July~2016 (ObsID-17779). IGR\,J17480--2446 is marked with a red circle.  }
\end{figure}

\subsection{IGR\,J17480--2446}
\label{subsec:p4:x2}

IGR\,J17480--2446 (here onwards J1748) is a LMXB that harbours an 11~Hz accreting X-ray pulsar primary, and a secondary that has an estimated mass of $\simeq$0.8-1.0 M$_\odot$   \citep{2010ATel.2929....1S,strohmayer10,testa12,patruno2012}. The 11-Hz pulsations and the type-I X-ray bursts \citep{2010ATel.2929....1S} observed from this source established the neutron star nature of the primary star. The source is located in the globular cluster Terzan\,5 (Fig.\,\ref{fig:p4:fov}) at a distance of $D\simeq$ 5.5~kpc \citep{ortolani07}. J1748 exhibited an outburst of approximately two months in 2010\footnote{Currently the 2010 outburst is the only outburst identified from J1748 although some of the old outburst activity observed from the direction of Terzan 5 may have been due to J1748 as well. However, this remains uncertain as the spatial resolutions of the earlier X-ray missions that observed these outbursts do not allow us to resolve sources in dense globular cluster cores \citep[see also the discussion in][]{degenaar2012exo1745}. } \citep{degenaar2011Ter5-414}.

Before the 2010 outburst of J1748, Terzan 5 was the target of several {\it Chandra} observations during which no X-ray transient was active \citep[][]{wijnands2005,heinke2006}. Due to these observations, the pre-outburst quiescent counterpart of J1748 could be readily identified \citep[][]{pooley2010} which serves as a proxy for the quiescent base level of this transient \citep[and hence gives us a clear indication of the thermal state of the neutron star core in this system;][]{degenaar2011Ter5-412}. After the accretion outburst of 2010, \citet{degenaar2011Ter5-414} found (using a {\it Chandra} observation about two months after the end of this outburst) that the quiescent counterpart was significantly brighter than before this outburst. This strongly indicates that the crust of the neutron star in this source was heated considerably out of equilibrium with the core and this demonstrates that a short (only a few months) outburst can also cause significant heating of the crust (likely due to a strong shallow heat source present in these systems). Frequent monitoring of the source using about a dozen {\it Chandra} follow-up observations allowed the cooling of the neutron star crust to be studied \citep{degenaar2011exo0748,degenaar2013,degenaar2015ter5x3}. Here we report on our continuous monitoring of the source using additional {\it Chandra} observations obtained in the summer of 2016 to further investigate its crust-cooling behaviour. 

\section{Observations and Data Analysis}
\label{sec:p4:obs}

\begin{table}
	\centering
	\caption{Log of the {\em Chandra} observations of J1748 used in our study}
	\label{tab:p4:obs}
	\begin{tabular}{clcc} 
		\hline
ObsID & Date      &    	Exposure Time       & Count Rate$^{1}$\\
	  &           &         (ks)  	        & (10$^{-3}$~c\,s$^{-1}$) \\\hline
 3798 & 2003 Jul 13/14 & 30.8$^{2}$  & 0.87$\pm$0.17         \\
10059 & 2009 Jul 15/16 & 36.3  & 1.17$\pm$0.18            \\
13225 & 2011 Feb 17    & 29.7 & 6.18$\pm$0.46           \\
13252 & 2011 Apr 29    & 39.5 & 3.75$\pm$0.31          \\
13705 & 2011 Sep 5     & 13.9 & 3.37$\pm$0.49           \\
14339 & 2011 Sep 8     & 34.1 & 3.07$\pm$0.30           \\
13706 & 2012 May 13    & 46.5 & 2.61$\pm$0.24           \\
14475 & 2012 Sep 17/18 & 30.5  & 2.90$\pm$0.31          \\
14476 & 2012 Oct 28    & 28.6  & 2.40$\pm$0.29           \\
14477 & 2013 Feb 5     & 28.6  & 1.80$\pm$0.25          \\
14625 & 2013 Feb 22    & 49.2  & 2.12$\pm$0.21           \\
15615 & 2013 Feb 23    & 84.2 & 1.96$\pm$0.15           \\
14478 & 2013 Jul 16/17 & 28.6  & 2.16$\pm$0.28           \\
14479 & 2014 Jul 15    & 28.6  & 1.73$\pm$0.25           \\
16638 & 2014 Jul 17    & 71.6  & 1.82$\pm$0.16           \\
15750 & 2014 Jul 20    & 23.0 & 2.12$\pm$0.30          \\
17779 & 2016 Jul 13/14 &      68.8 & 1.31$\pm$0.14           \\
18881 & 2016 Jul 15/16 &      64.7 & 1.25$\pm$0.14          \\ \hline
\multicolumn{4}{l}{1: Net (background subtracted) count rate (with 1$\sigma$ errors) }\\
\multicolumn{4}{l}{of J1748 in the energy range 0.5--10~keV. Typical (0.5-10 keV)}\\
\multicolumn{4}{l}{  background count rates during the observations are}\\
\multicolumn{4}{l}{  $\sim2- 8 \times 10^{-5}$ counts s$^{-1}$. }\\
\multicolumn{4}{l}{2: After removal of background flares.}\\
	\end{tabular}
\end{table}

We make use of archival {\em Chandra} observations\footnote{Obtained from \href{http://cxc.harvard.edu/cda/}{http://cxc.harvard.edu/cda/}} of Terzan 5 from 2003 to 2014 \citep[this same set of observations was used by us before in our previous cooling studies of J1748;][]{degenaar2011Ter5-414,degenaar2011exo0748,degenaar2013,degenaar2015ter5x3}, combined with two new observations obtained in July 2016 (PI: Degenaar; see Table\,\ref{tab:p4:obs} for the log of the observations used in our study). The two observations from 2003 and 2009 were performed before the 2010 outburst of J1748 and were used to estimate the pre-outburst temperature of the neutron star, whereas the remaining observations were used to model the thermal evolution of the neutron star after its 2010 outburst. All observations were taken in the Faint, timed exposure mode, using the Advanced CCD Imaging Spectrometer S array
\citep[ACIS-S;][]{garmire2003}, with a single frame exposure time of 3.0~s for all but six observations (ObsID 10059 - 1.6~s; ObsIDs 144[75-79] - 0.8~s). The cluster was positioned on the S3 chip.

\begin{figure}
\includegraphics[clip=,width=0.7\columnwidth,angle=270]{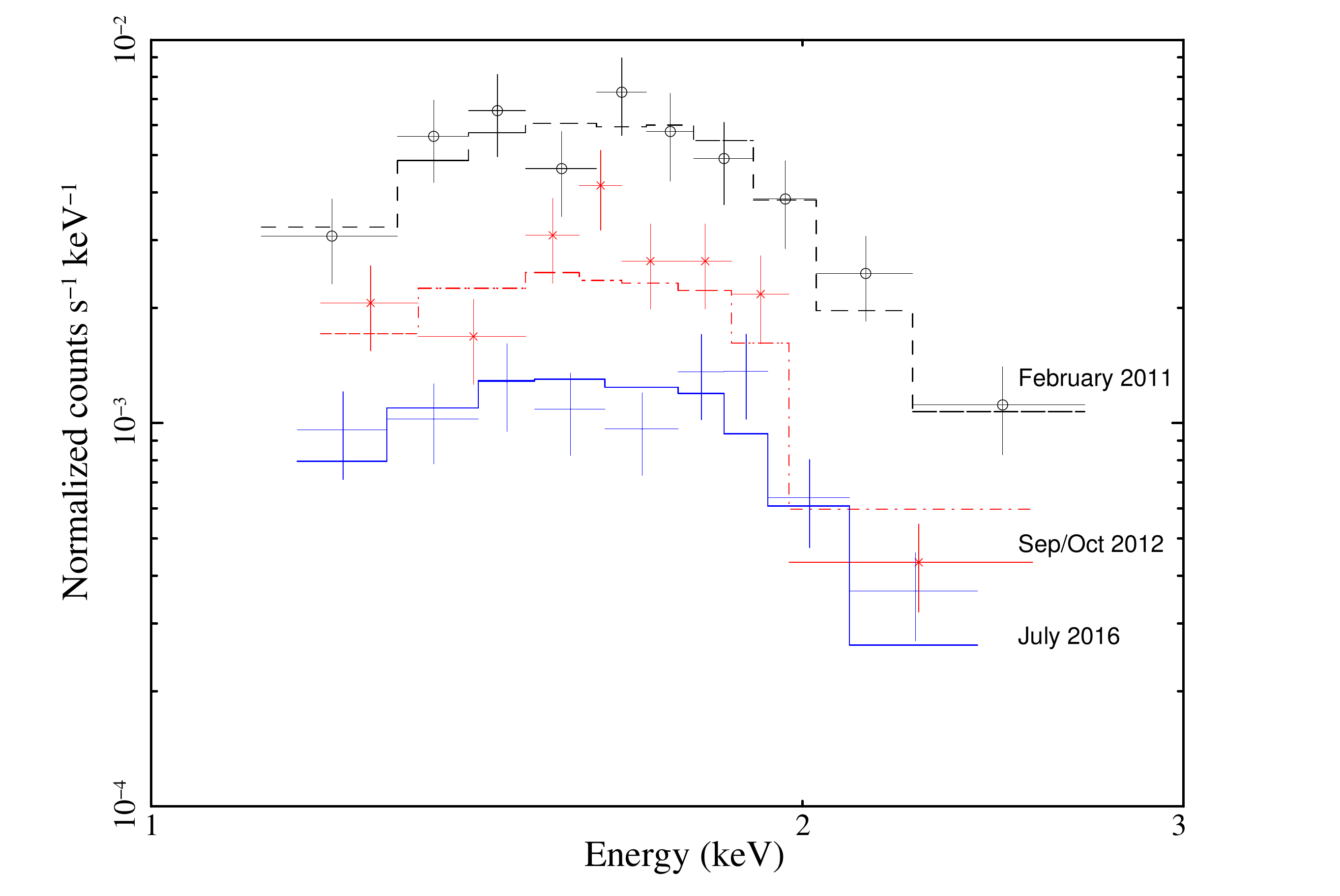}
\caption{The X-ray spectra obtained during different epochs of our {\em Chandra} observations -- the black points ($\circ$) represent the observation performed on February 2011 (ObsID 13225); the red points ($\times$) represent the combined spectra obtained during the observations obtained in September and October 2012 (ObsIDs 14475 and 14476); the blue points ($+$) represent the combined spectra from the data obtained during the observations performed in July 2016 (ObsIDs 17779 and 18881). The individual best-fit \textsc{nsatmos} models for each spectrum are represented by similarly coloured lines (dashed for Feb 2011, dash-dot-dash for Sep/Oct 2012, solid lines for Jul 2016). We only show three spectra, from the highest observed count rate (February 2011) to the lowest count rate (July 2016) detected so far, in order to avoid overcrowding in the figure. }
\end{figure}
\label{fig:p4:spectra}

We processed the level-1 event files using CIAO\footnote{\href{http://cxc.harvard.edu/ciao/}{http://cxc.harvard.edu/ciao/}} 4.8, and using the latest calibration files (CALDB\footnote{\href{http://cxc.harvard.edu/caldb/}{http://cxc.harvard.edu/caldb/}} 4.7.6). We created the level-2 event files with the \verb chandra_repro  script. To look for background flares, we created a background light curve for each observation in the energy range 0.5--10~keV using the \verb dmextract  routine. We found that during the observation performed on July 2003 (ObsID 3798) enhanced background flaring activity was present, showing more than 3$\sigma$ excursions from the average background count rate. The original exposure time for this observation was $\sim$39.3~ks and the final exposure time after removal of the flares was  $\sim$30.8~ks. The remainder of the observations were free of background flares. We obtained the count rates for J1748 during each observation using \verb dmextract \, (see Table\,\ref{tab:p4:obs}). We used a source extraction region of radius 1 arsec centered on the brightest pixel of the source in the X-ray image. For the background extraction region we made a selection on a source free part of the chip with a radius of 15 arcsec (this region was outside the field-of-view as shown in Figure \ref{fig:p4:fov} and therefore not shown in that figure). We used these extraction regions to obtain the source and background spectra; the corresponding response files were generated using \verb specextract .

\begin{table*}
	\centering
	\caption{The results obtained from our X-ray spectral analysis}
	\label{tab:p4:specfit}
	\begin{tabular}{lcccccc} 
		\hline
Epoch       &     MJD   & $kT_\text{eff}^{\infty}$& $F_\text{X}$ & $F_\text{bol}$ & $L_\text{X}$ & $L_\text{bol}$\\
		    &           &   (eV)  	            & (10$^{-13}$ erg\,cm$^{-2}$\,s$^{-1}$)	&(10$^{-13}$ erg\,cm$^{-2}$\,s$^{-1}$)	&  (10$^{33}$ erg\,s$^{-1}$) & (10$^{33}$ erg\,s$^{-1}$)  \\\hline
2003/2009	 &  52833.6/55027.7&	77.7$\pm2.0$ &	1.3$\pm$0.2	&	1.8$\pm$0.3        &	0.5$\pm$0.1	&	0.7$\pm$0.1	\\ 
Feb 2011     &	55609.4 &	107.6$\pm$1.8       &	6.0$\pm$0.5 &	7.4$\pm$0.6        &	2.2$\pm$0.2 &	2.7$\pm$0.2 \\
Apr 2011     &	55680.7 &	98.2$\pm1.6$ &	4.0$\pm$0.4 &	5.1$\pm$0.5        &	1.5$\pm$0.1 &	1.8$\pm$0.2 \\
Sep 2011     &	55812.7 &	96.5$\pm1.6$ &	3.7$\pm$0.3 &	4.7$\pm$0.4        &	1.3$\pm$0.1 &	1.7$\pm$0.2 \\
May 2012     &	56060.8 &	90.9$\pm$1.7        &	2.8$\pm$0.3 &	3.7$\pm$0.4        &	1.0$\pm$0.1 &	1.3$\pm$0.1 \\
Sep/Oct 2012 &	56207.9 &	92.5$\pm$1.6        &	3.1$\pm$0.3 &	3.9$\pm$0.4        &	1.1$\pm$0.1 &	1.4$\pm$0.1 \\
Feb 2013     &	56336.8 &	89.0$\pm1.2$ &	2.5$\pm$0.2 &	3.3$\pm$0.2        &	0.9$\pm$0.1 &	1.2$\pm$0.1 \\
Jul 2013     &	56489.9 &	91.0$\pm2.3$ &	2.8$\pm$0.4 &	3.7$_{-0.5}^{+0.6}$ &	1.0$\pm$0.2 &	1.3$\pm$0.2 \\
Jul 2014     &	56855.8 &	88.0$\pm1.5$ &	2.4$\pm$0.2 &	3.1$\pm$0.3        &	0.9$\pm$0.1 &	1.1$\pm$0.1 \\
Jul 2016     &	57583.6 &	84.8$\pm1.4$        &	2.0$\pm$0.2 &	2.7$\pm$0.2        &	0.7$\pm$0.1 &	1.0$\pm$0.1 \\ \hline
	\end{tabular}
\raggedright
Notes: The combined \textsc{nsatmos} fit using all spectra resulted in a neutral hydrogen column density, $N_\text{H}$, of 2.96$^{+0.12}_{-0.11}\times10^{22}$ cm$^{-2}$. This value is slightly higher than what was reported by \citet{degenaar2015ter5x3}, presumably because we are using the latest {\it Chandra} calibration files. The reduced $\chi^{2}$ was 1.039 for 72 degrees of freedom. $F_\text{X}$ and $L_\text{X}$ are the unabsorbed X-ray flux and X-ray luminosity for 0.5--10~keV, and $F_\text{bol}$ and $L_\text{bol}$ are the unabsorbed bolometric flux and luminosity (using the 0.01--100~keV range as a proxy). The errors are at the 1$\sigma$ confidence levels.

\end{table*}

For observations that were taken close to each other in time, we combined the spectra using the CIAO tool \verb combine_spectra . The quiescence observation of 2003 and 2009 were also combined to increase the statistics of the pre-outburst spectrum. By doing this, we assume that the neutron star temperature was constant in the quiescent period preceding the 2010 outburst of the source. However, this appears to be a valid assumption since \citet{degenaar2011Ter5-412} found that the two individual spectra were consistent with each other. We grouped the obtained spectra using \verb GRPPHA  such that each bin contained a minimum of 15 photons, and then fitted (using $\chi^2$ minimization) the ten spectra we obtained in the 0.5--10 keV energy range using \verb XSPEC \footnote{\href{https://heasarc.gsfc.nasa.gov/xanadu/xspec/}{https://heasarc.gsfc.nasa.gov/xanadu/xspec/}}. We also fitted the data using the Cash statistics \citep[][i.e., the way it is implemented in XSPEC so that it can also be used on background subtracted spectra; the so-called W-statistics]{cash1979} and we found that all data points had consistent temperatures between the two fit methods. In addition, the evolution of the temperature in time (see Section \ref{sec:p4:results}) was very similar. To not complicate the paper, we only display the results obtained using the $\chi^2$ statistics.

\section{Results}
\label{sec:p4:results}

\subsection{Observational results}

We fitted all the spectra simultaneously using the neutron star atmosphere model (\textsc{nsatmos}; \citealt{heinke06}) in combination with \textsc{tbabs} to model the interstellar absorption (using the {\tt WILM} abundances and the {\tt VERN} cross-sections; \citealt{wilms00,verner96}). In the \textsc{nsatmos} model, we fixed the distance to 5.5~kpc \citep{ortolani07}, and initially set the neutron star mass and radius to ${M= 1.4~\text{ M}_\odot}$ and  ${R= 10\text{ km}}$, similar to what was used in our previous studies of this source \citep[][see Section \ref{sec:mr} for the effects on our results when using different assumed masses and radii]{degenaar2011Ter5-414,degenaar2011ter5-418,degenaar2013}. The neutron star was assumed to be emitting from its entire surface, which means that the normalisation was set to unity. The hydrogen column density ($N_\text{H}$) was assumed to remain constant through all the epochs. Hence, this parameter was tied between all the spectra. This model results in a satisfactory fit -- $\chi^{2}/\nu=1.039$ for 72 degrees of freedom ($\nu$). The unabsorbed fluxes (0.5--10~keV) and corresponding X-ray luminosities were obtained by including the convolution model \textsc{cflux}. We also calculate the 0.01--100~keV fluxes and luminosities of the neutron star at each epoch, which can serve as good indicators for the bolometric fluxes\footnote{ The cut-off energies (both at the low and high energy end) of the \textsc{nsatmos} model vary with the temperature. However, for the temperatures of interest in our paper, the model is defined down to at least 0.01 keV and the flux ignored below this energy is always <0.1\% of the total flux. At the upper bound, the model is defined up to at least several tens of keV (again depending on temperature) but with typically very low fluxes above this cut-off energy.} and luminosities during each observation. Finally, we obtained the effective neutron star temperature as seen by an observer at infinity, $kT_{\text{eff}}^{\infty}$ \footnote{$kT_{\text{eff}}^{\infty}=kT_{\text{eff}}/(1+z)=kT_{\text{eff}}\cdot(1-R_\text{s}/R)^{1/2}$. Here, $k$ is the Boltzmann constant. For our assumed neutron star of radius $R=10$~km and mass $M=$1.4~M$_\odot$, the gravitational redshift $(1+z)=1.31$. With $R_\text{s}=2GM/c^2$ the Schwarzschild radius, $G$ the gravitational constant, and $c$ the speed of light.}. The neutron star temperatures, unabsorbed X-ray fluxes, and corresponding luminosities are shown in Table\,\ref{tab:p4:specfit}. The quoted uncertainties are at the 1$\sigma$ level of confidence. 

The 2010 outburst of J1748 started around ${\text{MJD}=55478}$ \citep{bordas10,linares10}. The end of the outburst was not observed because the source was in a Sun-constrained window around the time J1748 returned to quiescence. \citet{degenaar2011Ter5-414} estimated the end of the outburst (and thus the starting date of the cooling phase) to be at $t_0=55556$. Using this starting date for the  cooling phase we could construct the cooling curve as shown in Fig.~\ref{fig:nscool}. Clearly, the source cooled significantly over the course of $\sim$5.5 years; from $\sim$108 eV at the time of the first crust-cooling observation (in February 2011) to $\sim$85 eV during our new observations (July 2016). The same trend was reported by \citet{degenaar2015ter5x3} but we have extended the curve further using our new data. The new data point has a lower temperature compared to the previous point (obtained in July 2014; see Table~\ref{tab:p4:specfit}) demonstrating that, at that time, the crust was not in equilibrium yet with the core.

It should be emphasized though, that the last two data points in Figure~\ref{fig:nscool} (corresponding to the July 2014 and July 2016 observations) have measured temperatures that are almost consistent with each other.  However, the temperature decrease is supported by the fact that the source count rates (see Table~\ref{tab:p4:obs}) also decreased significantly between the corresponding observations. Moreover, when combining the 2014 July data (ObsIDs: 14479, 16638, 15750) together as well as the 2016 July data (ObsIDs: 17779 and 18881) we obtain a source count rate for these combination of observations of ${1.84 \pm 0.12 \times 10^{-3}\text{ counts s}^{-1}}$ and ${1.30 \pm0.10 \times 10^{-3}\text{ counts s}^{-1}}$, respectively.  Therefore, the count rate decrease is significant at the $\sim$3.4$\sigma$ level and hence we consider the observed decrease in temperature robust. This strongly indicates that the crust has cooled further and that it was not yet back in equilibrium with the core in July 2014. We note that this can be confirmed with future {\it Chandra} observations of the source.

It is also clear from Fig.~\ref{fig:nscool} that the source has not yet reached the same temperature as what was observed before the 2010 outburst. However, it is unclear if the source should indeed reach the same level or if it could level-off at a higher temperature. These two possibilities will be investigated further in the next section (Section \ref{subsection:nscool}) in which we model the observed cooling curve with our \verb NSCool ~cooling code. 

\subsubsection{Variation in mass and radius}\label{sec:mr}
Initially, we assumed a neutron star mass ${M = 1.4\text{ M}_\odot}$ and radius $R = 10 \text{ km}$ for our spectral fits, consistent with previous observational analysis. Under the assumptions made in our {\sc NSCool} code (see Section \ref{subsection:nscool}), such a neutron star has a crust that is $\sim800$ meters thick and a surface gravity $g_\text{s}=2.43\times 10^{14}\text{ cm s}^{-2}$. However, the crustal thickness can differ significantly depending on the mass and radius of the star; more compact stars  (i.e. higher surface gravity), have smaller crust radii and vice versa. This can affect the results when trying to constrain the crustal parameters. But, the assumed mass and radius also affect the obtained temperatures from spectral fitting, and hence for consistency the assumed mass and radius should be the same for the spectral fits and {\sc NSCool }models. 

To be able to have the mass and radius as free parameters in our simulations, we reanalysed all the spectral data for nine different combinations of mass and radius ($M=[1.0, 1.4, 1.8]\text{ M}_\odot$, and $R=[9,12,15] \text{ km}$), resulting in surface gravities ${g_\text{s}=(0.66-4.61)\times 10^{14}\text{ cm s}^{-2}}$ (crust thickness ${\Delta R\sim0.4-2.9 \text{ km}}$, where the thickest crust corresponds to the lowest surface gravity; see Table \ref{tab:gs14deltaR}). The results of the spectral analysis for different $M-R$ combinations are presented in Table \ref{tab:app3-temps} of Appendix \ref{section:appMRanalysis}.  The results show that the temperatures vary $\sim15$\% at most over the range of $M-R$ combinations. Fig. \ref{fig:MR-temp-plane} shows the surfaces in $M-R$ space formed by the obtained temperatures for the first and last observation. 

\subsection{Modelling the thermal evolution with {\sc NSCool}} 
\label{subsection:nscool}

We modelled the temperature evolution of the neutron star crust in J1748 during outburst and in quiescence using our thermal evolution code {\sc NSCool} \citep{page2013,page2016,ootes2016,ootes2018}. During outburst, the neutron star crust is heated out of thermal equilibrium with the core by the deep crustal heating processes  which, in our model, are assumed to be directly correlated with the rate of mass accretion onto the neutron star. We apply the deep crustal heating model from Table A.3 of \citet{haensel2008}, which assumes an initial composition of pure $^{56}$Fe below the layer of accreted matter that has been processed by X-ray bursts, and results in a total heat release of $1.93$ MeV per accreted nucleon. However, as discussed in Section \ref{section:p4:intro}, it has been found from previous studies of several of the other crust-cooling sources, that a second heating process of unknown origin must be active during the outburst at shallow depths \citep[see e.g.][]{brown2009}. The shallow heating properties (its strength and depth) have been found to differ significantly between sources and even between different outbursts of the same source \citep[e.g.][]{deibel2015,parikh2017maxi,ootes2018}. In our model, we assume it to be directly related to the accretion rate as well, although the validity of this assumption is currently unclear (see, e.g., the discussion about this in \citealt{ootes2018}, but also the recent results of \citealt{parikh2019mxb}).

To model the temperature evolution of the crust during outburst, we used the time-dependent accretion rate method as described in \citet{ootes2016}. The accretion rate of J1748 was determined from the 2-20 keV MAXI monitoring data of the source \citep{matsuoka2009}\footnote{The MAXI light curve of J1748 can be found at \href{http://maxi.riken.jp/star\_data/J1748-248/J1748-248.html}{http://maxi.riken.jp/star\_data/J1748-248/J1748-248.html} and is also shown in Figure 1 of \citet{degenaar2011Ter5-414}.}. 
For the accretion rate calculations, it is required to convert the count rates to bolometric fluxes. We used a conversion factor of $1.29\times10^{-8}\text{ erg cm}^{-2}\text{ count}^{-1}$ and again assumed a distance of $5.5$ kpc to be consistent with the spectral fits. This conversion factor was obtained from the bolometric luminosities reported by \citet[][]{papitto2012}\footnote{\citet{papitto2012} reported on the 0.05--150 keV luminosities which are reasonably accurate estimates for the bolometric ones. They quoted the luminosities for an assumed distance of 5.9 kpc so we converted them to 5.5 kpc (which we use in our work).}. For each reported observation, the bolometric luminosity is converted back to bolometric flux, which is then compared to the MAXI count rates observed on the same day to calculate a conversion factor. The final conversion factor is obtained by averaging the conversion factors of the individual daily-averaged MAXI observations. The average outburst accretion rate that we find with this method is $\langle\dot{M}\rangle=0.11~\dot{M}_\text{Edd}$ (where ${\dot{M}_\text{Edd}=1.5\times10^{18} ~{\rm g}~{\rm s}^{-1}}$ is the Eddington accretion rate), which is consistent with the value estimated by \citet{degenaar2011Ter5-412}.

The observed temperature evolution during the crust cooling phase depends on the  heating during the accretion outburst, the properties of the crust, and the temperature of the core. The amount of heating depends on the mass accretion rate, and the strength ($Q_\text{sh}$) and depth ($\rho_\text{sh}$) of the shallow heating mechanism. For other sources the amount of shallow heating is typically found to be ${Q_\text{sh}=0-2 \text{ MeV nucleon}^{-1}}$ (see e.g. \citealt{degenaar2015ter5x3} and references therein, and \citealt{merritt2016,parikh20181rxs, ootes2018} for more recent results), although at least one source needs significantly more shallow heat during one outburst (${Q_\text{sh}=6-17\text{ MeV nucleon}^{-1}}$; \citealt{deibel2015,parikh2017maxi}). 

The parameters affecting the thermal conductivity (and thus the temperature evolution) are the composition of the envelope (the region with $\rho<10^8$ g~cm$^{-3}$, which acts as a heat blanket), parameterized by the amount of light elements in this layer, and the level of impurity in the solid part of the crust (set by the impurity parameter $Q_\text{imp}$). The envelope has a higher thermal conductivity if it consists of more light elements, which results in a higher surface temperature for the same internal temperature (i.e. the temperature below the envelope). A larger impurity factor decreases the thermal conductivity of the crust due to electron-impurity scattering. Within {\sc NSCool}, one can define multiple density regions with different impurity factors. Typically, the impurity parameter throughout the crust is found to be $Q_\text{imp}\sim1-10$ \citep[e.g.][]{brown2009,cumming2017} from the modelling of observed cooling curves of several sources, although the impurity of the pasta layer (the innermost crust layer) might be larger \citep[e.g.][]{horowitz2015,deibel2017}. 

The specific heat of the crust depends on the presence of neutron superfluidity in the inner crust. If the dripped neutrons form a superfluid, they do not contribute to the neutron specific heat, hence reducing the total crust specific heat. The core temperature (parameterized by the gravitationally redshifted core temperature before the onset of accretion, $\widetilde{T}_0$) sets the equilibrium temperature between the crust and the core and thus affects the base level of the cooling curve. The assumed mass and radius set the surface gravity, and with that the thickness of the crust.

\begin{figure}
\includegraphics[width=0.5\textwidth]{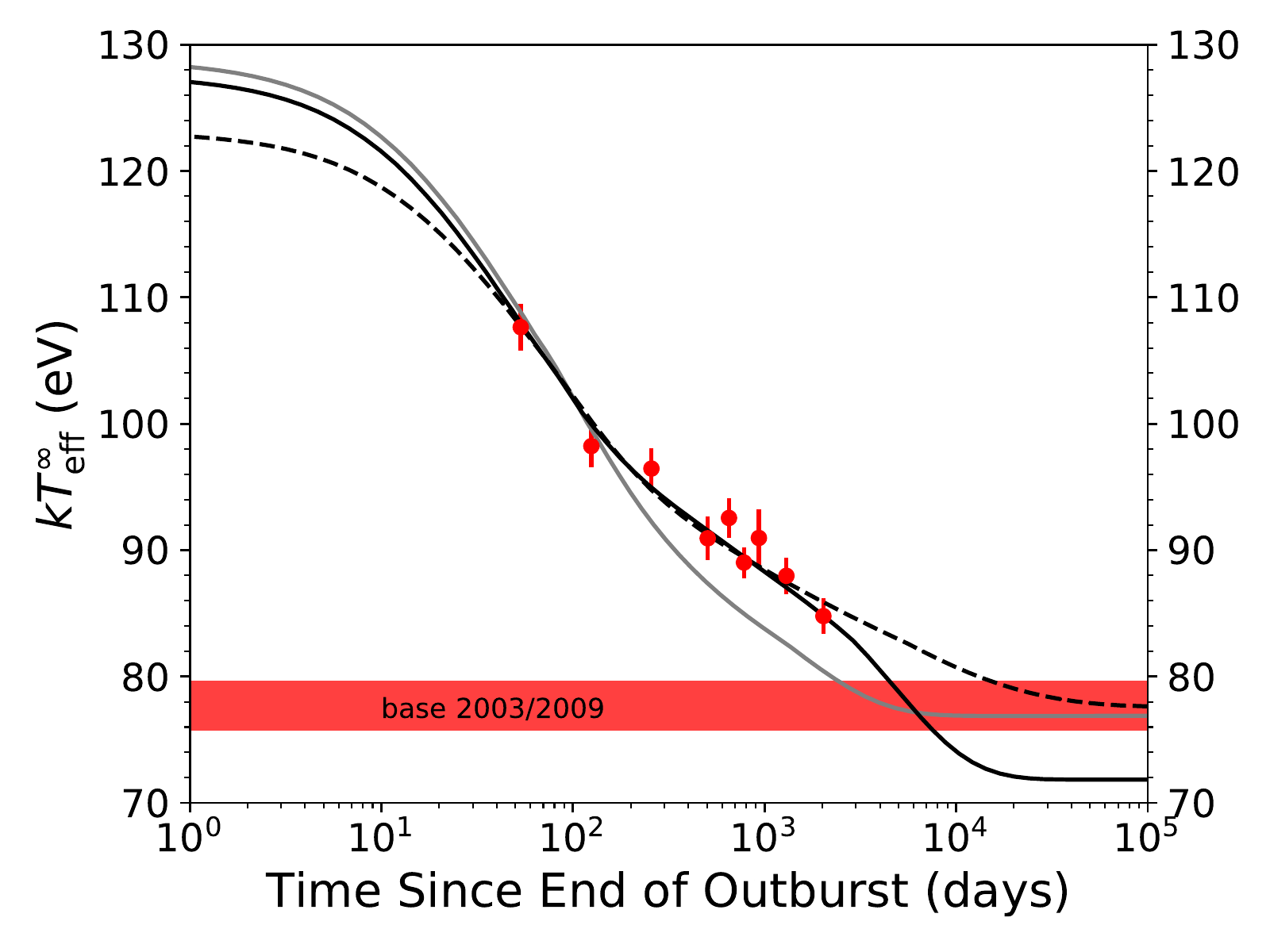}
\caption{ Observed cooling (red points) and exemplary cooling curves calculated using our {\sc NSCool} code, for assumed mass ${M=1.4\text{ M}_\odot}$ and radius ${R=10\text{ km}}$. The black solid curve serves as an example for model 1. This curve is not forced back to the observed pre-outburst base level (red horizontal band), and we assumed for this curve ${Q_\text{imp}^{(2)}=550}$, and ${Q_\text{imp}=25}$ for the shallower (${Q_\text{imp}^{(1)}}$) and deeper (${Q_\text{imp}^{(3,4,5)}}$) impurity zones in the crust. The increased impurity factor delays the cooling after $>100$~d. The effect is shown by the grey curve, which shows a model in which the same parameters are assumed as for the solid black curve, except ${Q_\text{imp}^{(2)}}$, which is also set to 25. Note that this curve is not forced to return the pre-outburst base level, the overlap of the grey curve with the red horizontal band is coincidental. The black dashed curve is an example curve for model 2, which is forced to cool down to the 2003/2009 base level. }\label{fig:nscool}
\end{figure}

\subsubsection{Exploring the parameter space with the Markov chain Monte Carlo method}\label{sec:mcmc}

The free parameters in our model are numerous (compared to the number of data points) and moreover degenerate, which makes it difficult to constrain the parameters of our model using simple fitting methods. Therefore, we did a full exploration of the parameter space in search for the parameters that best reproduce the data. This is done by using {\sc NSCool} together with our recently developed Markov chain Monte Carlo (MCMC) driver {\tt MXMX} \citep[see][for full details]{page2019}, which allows us to consider a larger number of parameters\footnote{MCMC with Bayesian statistics works even if the number of parameters is larger than the number of data points (as is the case in our studies) but will likely then show that most of these parameters are unconstrained.}.

{\tt MXMX} is inspired from {\tt emcee} \citep{Foreman-Mackey2013}, but written in Fortran and designed for maximum efficiency when coupled with {\sc NSCool}. Here we use it with 100 walkers and run it for 40 thousand steps (with auto-correlation times typically of the order of 50 to 250). In total, we thus run more than four million cooling curves and eliminate an initial set of points ("burn-in points") to keep only the part of the chain after convergence (see Fig. \ref{fig:AppMCchi2}). In all cases the number of points we use in our statistics is at least a hundred times more than the integrated auto-correlation time \citep{sokal1997} of the chain. The MCMC simulations show the $\chi^2$ distribution to have a second local minimum, i.e. the distribution is bi-modal, as displayed in Fig. \ref{fig:AppMCchi2bimodal}, but with a large ($> 75$) $\chi^2$, which requires the use of tempered chains in our simulations. Moreover, the time to reach convergence is long, which we determine by verifying that the $\chi^2$ distribution becomes stationary, as illustrated in Fig. \ref{fig:AppMCchi2}. 

We run two different models. The first model (model 1) fits the nine data points observed after the 2010 outburst ended. Since it is not clear from these observations wether the surface temperature of the neutron star has levelled off already (which would indicate that the source is back in crust-core thermal equilibrium) the base-level for this model is free. However, the source was also observed in quiescence before the start of its 2010 outburst. From observations taken in 2003 and 2009, a quiescence surface temperature of $77.7\pm2.0\text{ eV}$ was measured (assuming $M=1.4\text{ M}_\odot$ and $R=10\text{ km}$, see Table \ref{tab:p4:specfit}). Assuming that the source was in thermal equilibrium at the time of these pre-outburst observations, J1748 can be expected to eventually cool to this base level. Hence in the second model, we tested this scenario by forcing the calculated cooling curve to return to the 2003/2009 base level (the red horizontal band in Fig. \ref{fig:nscool}), by adding an additional point to the quiescence observations. This point, with a temperature equal to the one determined from the 2003/2009 observations, was set at a time of $10^5$ days after the end of the outburst, such that we leave the time at which the source returns to the pre-outburst base level (which is typically $\sim10^{3-4}$ days into quiescence) free. This requirement is not added to model 1, because the source may cool down to a new base level. Even though the core temperature $T_0$ is not expected to change considerably over the time of the 2010 outburst, the base level after the outburst can still be significantly different due to a change in envelope composition compared to that at the time of the pre-outburst measurement.

Our MCMC parameters, and their {\em prior} distributions, are as follows:
\begin{itemize}
\item The mass $M$, with a uniform prior distribution between $1.0\text{ M}_\odot$ and $1.8\text{ M}_\odot$.
\item The radius $R$, with a uniform prior distribution between $9.0$ and $15.0$ km.
\item Initial (redshifted) core temperature ($\widetilde{T}_0$), with a $\log \widetilde{T}_0$ prior distribution chosen as uniform between $7$ and $9$.
\footnote{
 Due to general relativistic effects, in a thermally relaxed state the local temperature $T$ is {\em not} uniform in the stellar interior, but it is rather the redshifted temperature $\widetilde{T} \equiv \mathrm{e}^\phi T$ which is uniform, the redshift factor $\mathrm{e}^\phi$ being given by the time part of the metric $g_{00}$ through $g_{00} = \mathrm{e}^{2\phi}$ (see, e.g., \citealt{page2004}).}
\item The impurity parameter ($Q_\mathrm{imp}$) within the crust is divided into 5 regions according to density: $Q_\text{imp}^{(1)}$ for the region ${\rho<10^{11}\text{ g cm}^{-3}}$, $Q_\text{imp}^{(2)}$ for ${\rho=10^{11-12}\text{ g cm}^{-3}}$, $Q_\text{imp}^{(3)}$ for ${\rho=10^{12-13}\text{ g cm}^{-3}}$, $Q_\text{imp}^{(4)}$ for ${\rho=10^{13-14}\text{ g cm}^{-3}}$, and $Q_\text{imp}^{(5)}$ for ${\rho>10^{14}\text{ g cm}^{-3}}$.
All $Q_\mathrm{imp}^{(i)}$ have prior distributions uniform between 0 and 1500. The upper bound allows for the possibility of highly amorphous crust structures.
\item The column densities of Helium ($y_\text{He}$) and Carbon ($y_\text{C}$) in the envelope, with values chosen from uniform distributions between $0$ and $10$ for $\log y_\text{He}$ and between $0$ and $12$ for $\log y_\text{C}$, both in units of g cm$^{-2}$.
\item Shallow heating strength $Q_\mathrm{sh}$, with uniform prior distribution between $0$ and $5$ MeV per accreted nucleon. 
\item The minimum density ${\rho_\mathrm{sh}^\mathrm{min}}$ and maximum density $\rho_\mathrm{sh}^\mathrm{max}$ of the shallow heating layer where the energy $Q_\mathrm{sh}$ is injected. Values of $\log \rho_\mathrm{sh}^\mathrm{min}$ and $\log \rho_\mathrm{sh}^\mathrm{max}$ are chosen from a uniform distribution between $8$ and $\log \rho_\text{cc}$, in units of $\text{g cm}^{-3}$. Here, $\rho_\text{cc}$ is the crust-core transition density, which we fix at ${1.5 \times 10^{14}\text{ g cm}^{-3}}$. Within the layer ranging between densities $\rho_\mathrm{sh}^\mathrm{min}$ and $\rho_\mathrm{sh}^\mathrm{max}$, the $Q_\mathrm{sh}$ heat is uniformly injected per unit volume.

\item The fraction ($a_\mathrm{entr}$) of dripped neutrons in the inner crust which are entrained by the nuclei in their thermal motion, uniformly distributed between $0$ and $1$.
This parameter increases the nuclei effective mass and affects their specific heat \citep{Chamel2013} as well as the electron-phonon scattering rate a low temperature (but the latter is actually ineffective here because of impurity scattering that dominates unless $Q_\mathrm{imp}$ is very small). In a similar fashion as $Q_\mathrm{imp}$, $a_\mathrm{entr}$ should be considered as density dependent, but in this study we adopt a single value in the whole inner crust and,
as we will see, its effect is negligible.
\item The density range, $\rho_\mathrm{SF}^\mathrm{min}-\rho_\mathrm{SF}^\mathrm{max}$, in which dripped neutrons in the inner crust are assumed to be superfluid with a high critical temperature, ${T_c=10^{10}\text{ K}}$, while they are normal below and above this range. 
The values of $\rho_\mathrm{SF}^\mathrm{min}$ and $\rho_\mathrm{SF}^\mathrm{max}$ are constrained to be larger than the neutron drip density ${\rho_\text{drip} \simeq 7 \times 10^{11} \text{ g cm}^{-3}}$ and smaller than $\rho_\text{cc}$.
Both the $\log \rho_\mathrm{SF}^\mathrm{min}$ and $\log \rho_\mathrm{SF}^\mathrm{max}$ values are chosen from uniform distributions within their allowed ranges. The effect of superfluidity is to strongly suppress the neutron specific heat within this density range.
\end{itemize}
 
Our $M$ and $R$ parameters are chosen within the range of our $M-R$ grid assumed for the data analysis. As mentioned before, one should, for consistency, assume the same mass and radius for the model simulation as was assumed for the analysis of the observations. Therefore, for each point in our MCMC simulation, we interpolate the temperatures from Table \ref{tab:app3-temps} in accordance with the $M$ and $R$ assumed in that calculated cooling curve.

\begin{figure*}
\includegraphics[width=0.9\textwidth]{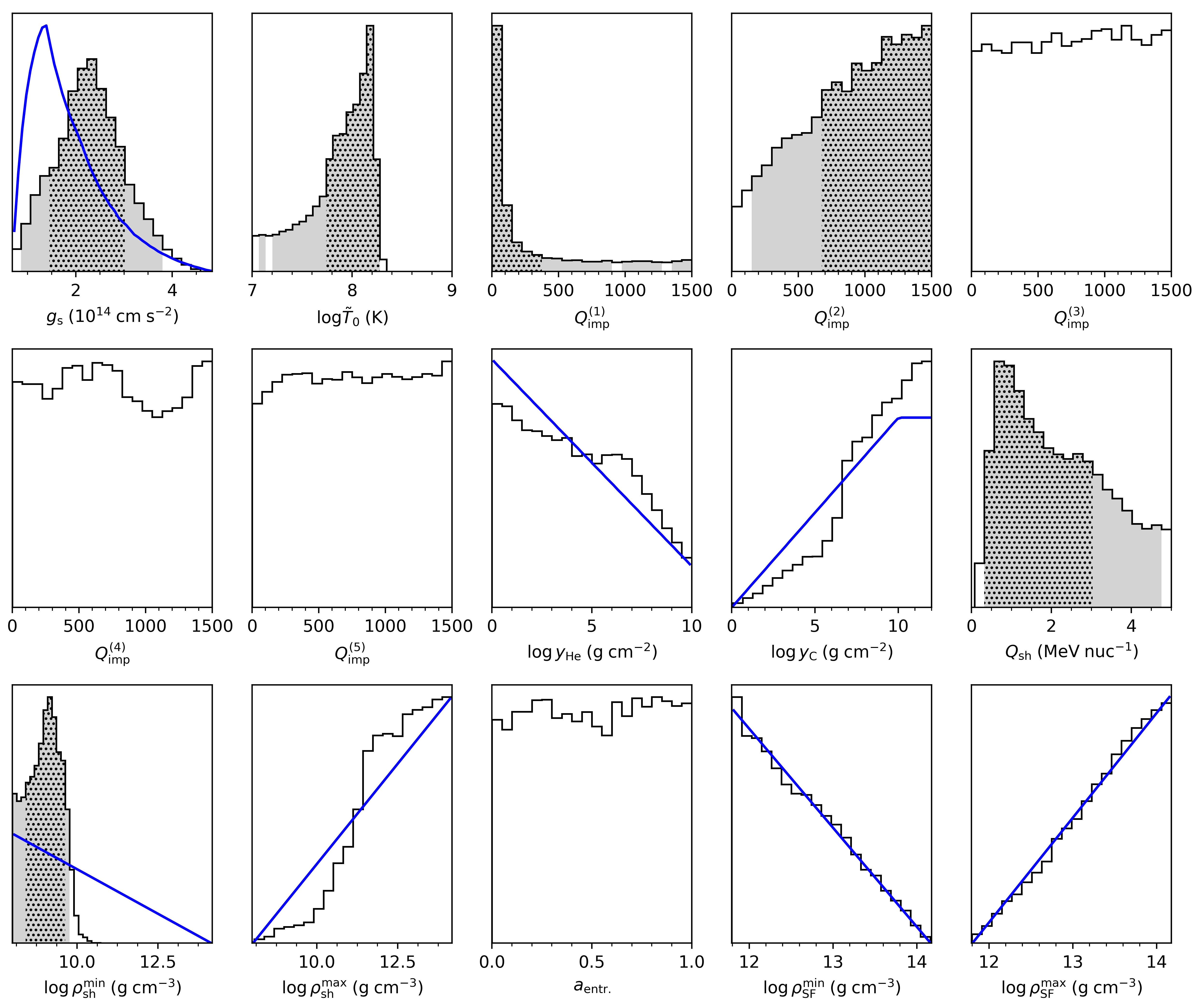}
\caption{Histograms of the posterior probability distributions of our MCMC parameters for model 1. We show here $g_\text{s}$ instead of $M$ and $R$, and plot for the parameters that have non-uniform prior distributions (i.e. $g_\text{s}$, $\log{y_\text{He}}$, $\log{y_\text{C}}$, $\log{\rho_\text{sh}^\text{min}}$, $\log{\rho_\text{sh}^\text{max}}$, $\log{\rho_\text{SF}^\text{min}}$, and $\log{\rho_\text{SF}^\text{max}}$) their prior distribution in blue for comparison. The grey-filled bins indicate the region of decreasing likelihood that makes up 95\% of the area under the distribution and similarly, the dot-filled bins indicate the region that makes up 68\% of the total area. These areas are omitted if the posterior distribution is (nearly) identical to the prior distribution, indicating that the parameter cannot be constrained from this model. Vertical scale is arbitrary. See text for details.}
\label{fig:histomodel1}
\end{figure*}

Although all values are chosen from uniform distributions, we additionally require all values to be physical. This means that we require $y_\text{C}>y_\text{He}$, $\rho_\text{sh}^\text{max}>\rho_\text{SF}^\text{min}$, and $\rho_\text{SF}^\text{max}>\rho_\text{SF}^\text{min}$. This leads to non-uniform prior distributions for these six parameters (see Fig. \ref{fig:histomodel1} for the prior distributions of the parameters for which the prior is not uniform).

The total number of free parameters in these simulations is larger than the number of data points. However, each one of them controls an important physical (heating rate, thermal conductivity and specific heat) or geometrical (curst thickness through $g_\mathrm{s}$ by $M$ and $R$) ingredient of our cooling simulations. Most of these parameters will be unconstrained by the data but instead of choosing {\em a priori} which ones are or are not important we prefer to keep maximum freedom and let the MCMC method guide us.

\begin{figure*}
\includegraphics[width=0.99\textwidth]{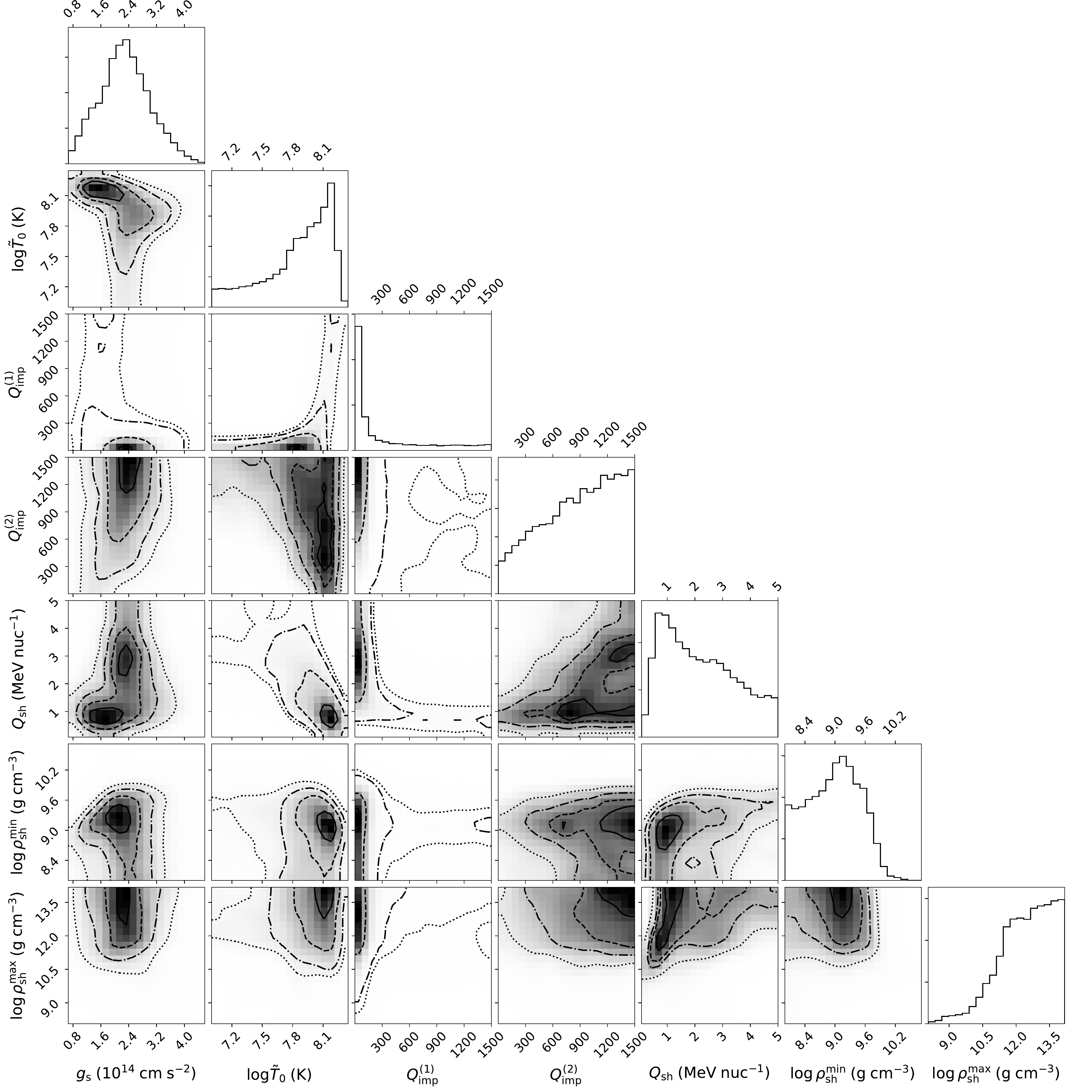}
\caption{Results of our model 1 MCMC for selected parameters: $g_\text{s}$, $\tilde{T}_0$, $Q_\text{imp}^{(1)}$ and $Q_\text{imp}^{(2)}$ (i.e. the impurity factor for the outer two crust layers $\rho<10^{11}$ and $\rho=10^{11-12}\text{ g cm}^{-3}$), and shallow heating parameters (strength and depth). The solid, dashed, dashed-dotted, and dotted curves show respectively, the 0.5, 1.0, 1.5, and 2.0$\sigma$ levels in the 2D histograms.}
\label{fig:cornermodel1}
\end{figure*}

\subsubsection{Model 1: free base level}
 
The results of our MCMC run for model 1 are presented in Fig.~\ref{fig:histomodel1} and Fig.~\ref{fig:cornermodel1}. Fig. \ref{fig:histomodel1} shows the posterior distributions of all parameters in our MCMC run (but with $M$ and $R$ converted to $g_\text{s}$). In Fig. \ref{fig:cornermodel1} we present the 2D histograms of a selection of the parameters, which shows the relations between different parameters. In Fig. \ref{fig:nscool} we show the observed temperatures and some exemplary calculated cooling curves. The posterior distributions in Fig.~\ref{fig:histomodel1} show that, as expected, only a few of our MCMC parameters are constrained and that even for those the allowed parameter ranges are large. Nevertheless, we describe here the posterior distributions in attempt to compare them with expected values.

The posterior distribution of the surface gravity peaks around $\sim2.3\text{ cm s}^{-2}$. This peak includes the surface gravity of $g_\text{s}=2.43\text{ cm s}^{-2}$ that was assumed in our initial analysis of the observations (i.e. mass $M=1.4\text{ M}_\odot$ and radius $R=10\text{ km}$). However, the distribution also shows that only the most extreme surface gravities are outside the preferred range.

From our model 1, the core temperature cannot be constrained. The posterior distribution shows a preference for high core temperatures (with a peak around $\log\tilde{T}_0\sim8.1$~K), but lower values are still allowed. Only the highest values (${\log\tilde{T}_0\gtrsim8.3}$~K) can be excluded based on this simulation. 

The properties of the deepest layers of the crust remain unconstrained for this model as well, as the impurity parameter in these layers (layers 3--5, which coincide with densities ${\rho>10^{12}\text{ g cm}^{-3}}$), as well as the parameters for the superfluidity in the inner crust, are found to have posterior distributions that are very close to the corresponding priors. Additionally, we do not obtain any constraints on the dripped neutron entrainment, $a_\text{entr}$, which only affects the specific heat in the deep inner crust. This indicates that this star has to evolve further in time for its surface temperature to be sensitive to the physics of the deep inner crust (see the discussion of this $T_\text{eff}(t) - T(\rho)$ mapping in \citealt{brown2009}). Furthermore, no constraints on the envelope composition of J1748 can be obtained from model 1. The posterior distributions for $\log y_\text{He}$ and $\log y_\text{C}$ are close to the prior distributions.

On the other hand, the posterior distributions of the impurity parameter for the two outermost layers of the crust show a very interesting trend. Even though a precise value of $Q_\text{imp}^{(1)}$ and $Q_\text{imp}^{(2)}$ cannot be constrained from the posterior distributions in Fig.~\ref{fig:histomodel1}, the lowest values of the allowed range seem to be preferred for the impurity in the outermost region of the crust (impurity region 1; $\rho<10^{11}\text{ g cm}^{-3}$, although it should be noted that the grey shaded area in the panel for $Q_\text{imp}^{(1)}$ in Fig. \ref{fig:histomodel1} shows that larger values are not excluded), while in the second region of the crust, around the neutron drip density ($\rho=10^{11-12}\text{ g cm}^{-3}$), the posterior distribution shows a preference for high values of the impurity parameter. The 95\% confidence range in the panel for $Q_\text{imp}^{(2)}$ in Fig. \ref{fig:histomodel1}, shows a preference for values $Q_\text{imp}^{(2)}\gtrsim150$. Such a value is $\gtrsim10$ times larger than what has been found for other crust-cooling sources. The black curve in Fig. \ref{fig:nscool} represents a model with $Q_\text{imp}^{(2)}=550$, while the value in the other layers is lower ($Q_\text{imp}^{(1,3,4,5)}=25$). To show the effect of the high impurity layer for this particular curve, a grey curve is added to the figure which shows a cooling curve that assumes the same parameters as the black curve, but with a constant $Q_\text{imp}=25$ throughout the crust. In \citet{wijngaarden2018} we presented a preliminary model (using {\sc NSCool} as well) of the thermal evolution of J1748, but without our newly added data point from July 2016. In that work a different method was applied, which did not include the same free parameters. Nevertheless, we found a large impurity parameter for part of the crust in that work as well. 

\begin{figure*}
\centering\includegraphics[width=0.9\textwidth]{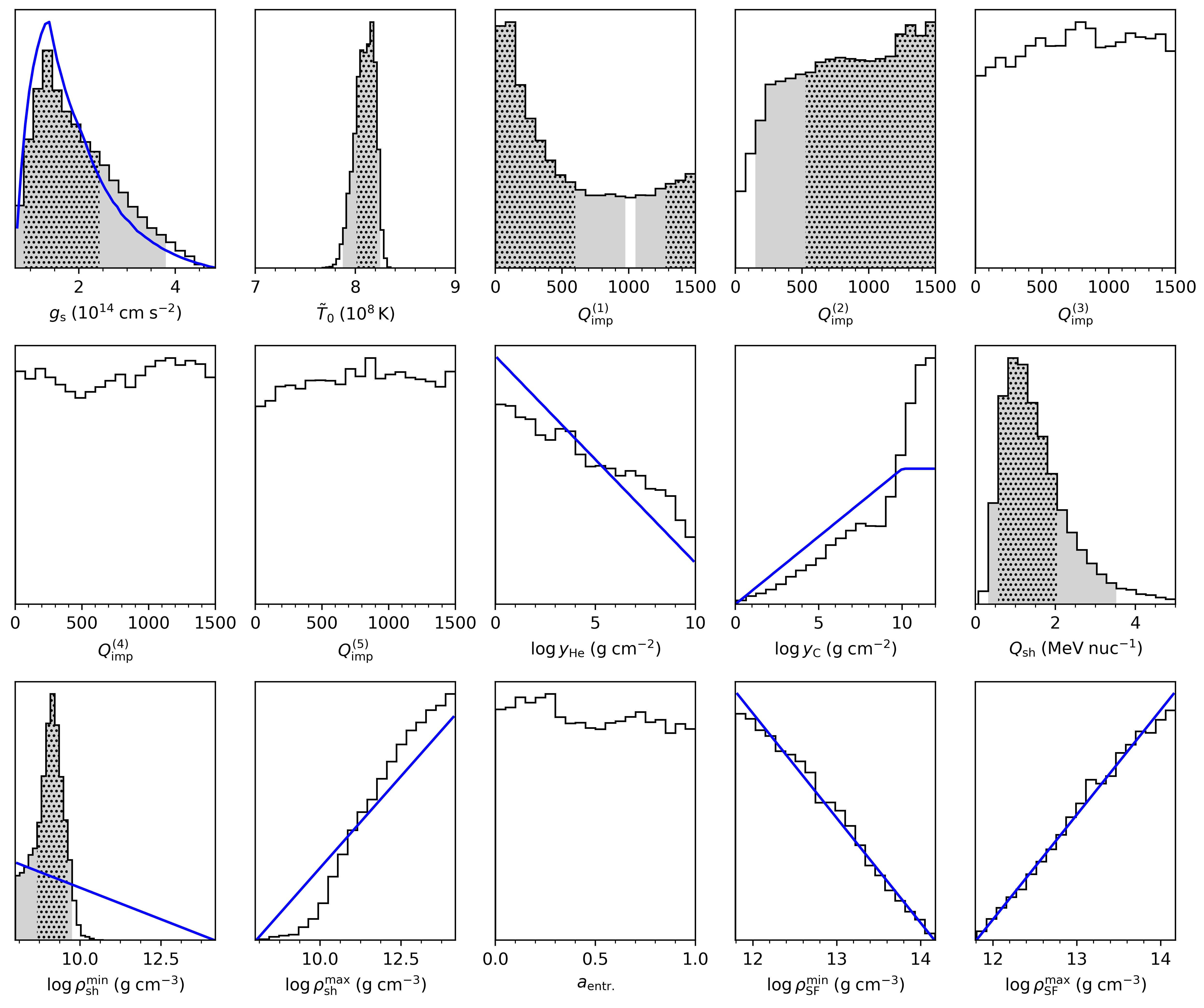}
\caption{Same is Fig. \ref{fig:histomodel1}, but for model 2 in which we force the source to cool down to the pre-outburst base level.}
\label{fig:AppMChistomodel2}
\end{figure*}

The shallow heating is modelled with three parameters. The posterior distribution of model 1 for the shallow heating strength peaks around ${\sim1\text{ MeV nucleon}^{-1}}$, and shows that to fit the cooling data, at least some shallow heating is required. We find a preference for a minimum depth of the shallow heating around $\sim10^9\text{ g cm}^{-3}$ (and with an allowed range up to $\sim10^{10}\text{ g cm}^{-3}$), in agreement with typical values found for other sources. The poster distribution for the maximum depth of the shallow heating layer allows for a large range of parameters, and shows that the shallow heating range may extend into the inner crust. However, our current observations do not allow us to constrain the inner crust yet. It is interesting to note that Fig. \ref{fig:cornermodel1} shows that the highest values of the shallow heating strength ($Q_\text{sh}\gtrsim3.5\text{ MeV nucleon}^{-1}$) are only allowed for large values of the maximum shallow heating depth ($\log\rho_\text{sh,max}\gtrsim 12\text{ g cm}^{-3}$), in which case most of the heat is deposited in the unprobed region.


\subsubsection{Model 2: including the pre-outburst base level}

\begin{figure*}
\includegraphics[width=0.99\textwidth]{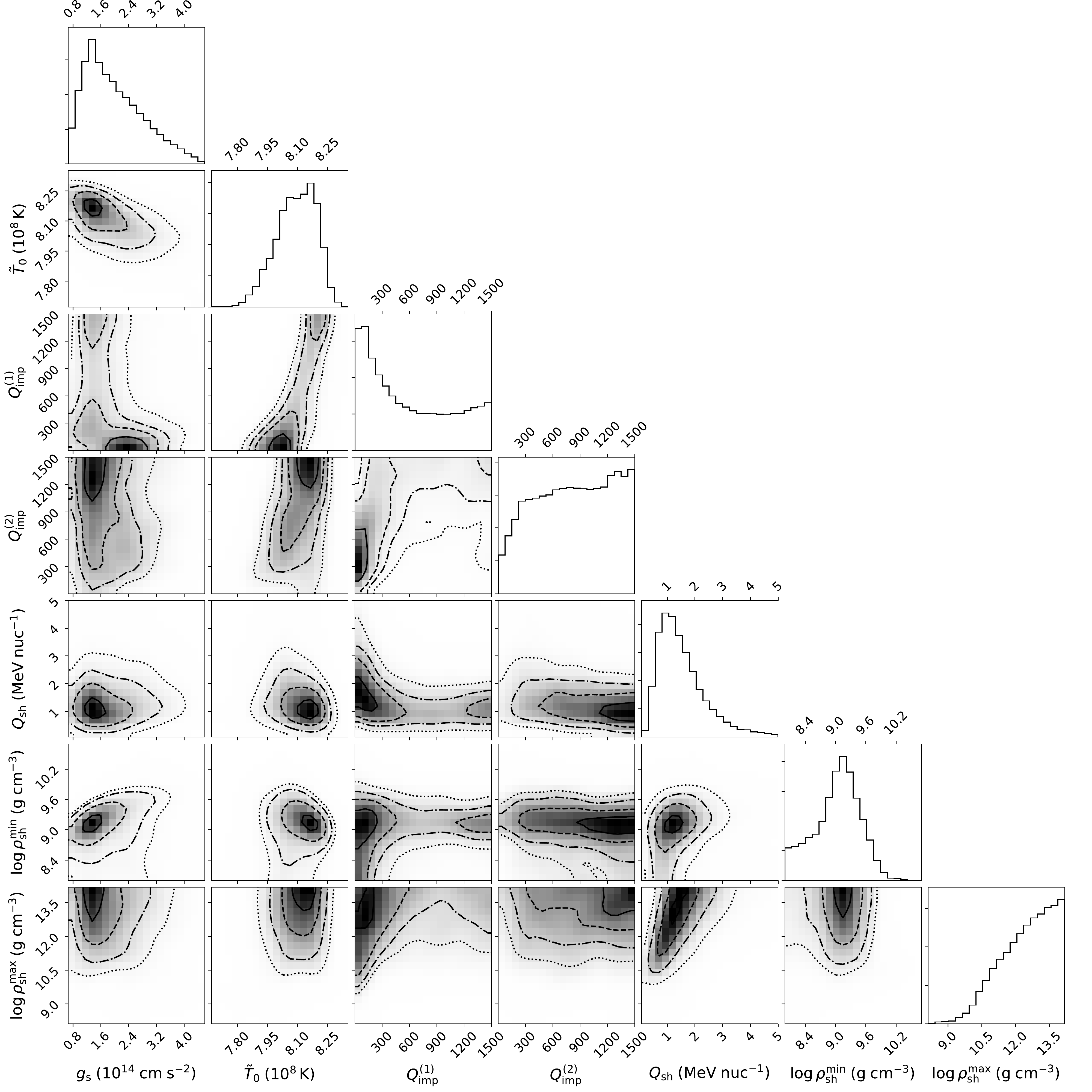}
\caption{Results of model 2 for a selection of parameters. The solid, dashed, dashed-dotted and dotted curves show respectively, the 0.5, 1.0, 1.5, and 2.0$\sigma$ levels in the 2D histograms.}
\label{fig:AppMCcornermodel2}
\end{figure*}

In model 2, a hypothetical observation is added to the cooling data to force further cooling towards to pre-outburst base level. The time at which base-level is reached is left free by adding this additional point at $10^5$ days after the end of the outburst. The results are shown in Fig. \ref{fig:AppMChistomodel2} and Fig. \ref{fig:AppMCcornermodel2}. 

As expected, model 2 places more constraints on the core temperature than model 1. The posterior distribution shows again a peak around ${\log\tilde{T}_0\sim8.1}$~K, but the allowed range is much more constrained (the grey area comprises ${\log\tilde{T}_0\sim(7.85-8.25)}$~K). The surface gravity and envelope composition are not constrained for model 2, and hence no conclusions can be drawn on the values of these parameters for J1748. 

The results for $Q_\text{imp}$ and the envelope composition for model 2 are similar to model 1. As a consequence of the constraint on the core temperature, the impurity parameter for the outmost region of the crust $Q_\text{imp}^{(1)}$ becomes less constrained compared to model 1 (see Fig.\ref{fig:AppMCcornermodel2}). We do find again that the impurity parameter in the second layer of the crust seems to prefer higher values ($Q_\text{imp}^{(2)}\gtrsim150$, from the grey area in Fig. \ref{fig:AppMChistomodel2}) than found for other sources. When the source is forced to cool down to the pre-outburst base level, the posterior distribution for the shallow heating strength peaks stronger around $\sim1\text{ MeV nucleon}^{-1}$ compared to model 1, but the overall results for the shallow heating parameters are the same.

A constraint on the deepest region 5, where the nuclear pasta is expected to be found with a likely large $Q_\text{imp}$, would require several data points in the future as was recently exemplified by the last observation of MXB 1659-29 and its interpretation \citep{deibel2017}. With multiple future observation the base level as well as the time at which crust-core equilibrium is reached can be constrained, which would allow for constraints on the physics of the deep crust. These considerations also explain the absence of any constraints from model 2, on the one hand, on the dripped neutron entrainment, $a_\text{entr}$, which only affect specific heat in the deep inner crust and, on the other hand, on the crust superfluidity.

\section{Discussion}
\label{sec:p4:discussions}

We used archival {\em Chandra} observations in conjunction with two new observations of Terzan\,5 to study the thermal evolution of the neutron star in the X-ray transient IGR\,J17480--2446. Previous observations (up to the summer of 2014) showed that the neutron star crust in this system was significantly heated (i.e., out of thermal equilibrium with the core) during the 2010 outburst of the source and in the sub-sequent quiescent period the crust had cooled significantly \citep{degenaar2011Ter5-412,degenaar2011Ter5-414,degenaar2011exo0748,degenaar2013,degenaar2015ter5x3}. The last observations reported previously were performed in July 2014 \citep{degenaar2015ter5x3}. In our paper we have presented new observations obtained two years later, in July 2016. With these new observations we span a post-outburst quiescent period of $\sim$5.5 years allowing us to study the crust-cooling behaviour of this source in more detail. We fitted the source spectrum with an absorbed neutron star atmosphere model (i.e., \textsc{nsatmos}) and found that the measured surface temperature was lower than the temperature observed during the July 2014 observations (see Table~\ref{tab:p4:specfit}).

By modelling the thermal evolution of J1748 with our crust heating and cooling code {\sc NSCool} \citep{page2013,page2016,ootes2016,ootes2018}, we have tried to constrain the crustal parameters of the neutron star in this source. We explore the full parameter space using MCMC simulations. The source has been modelled before by \citet{degenaar2011ter5-418,degenaar2013}, using a different code. In our current paper, we make use of more observations and use a more advanced code which takes into account accretion rate fluctuations based on the observed light curve \citep{ootes2016}. Taking into account variations in accretion rate is especially important for short-duration transients such as J1748, because it is further off from a steady-state temperature profile during the outburst. 

From our model 1 (which does not require the source to return to the 2003/2009 base level), we found that a shallow heating strength of the order of ${\sim1\text{ MeV nucleon}^{-1}}$ is required to explain the temperature evolution in quiescence. The amount of shallow heating is not strongly constrained, but the model shows that at least some shallow heating is needed to reproduce the observations. For the depth of the shallow heating we find a preference for a minimum depth around $\sim10^9\text{ g cm}^{-3}$, but full extent of the shallow heating region can not be constrained from this model. \citet{degenaar2013}, fitted the temperature data obtained using the initial observations performed between 2011 and 2013 with $Q_\text{sh}=1\text{ MeV nucleon}^{-1}$, similar to our results. On the other hand, \citet{turlione2015}, found that the data was best fitted when an additional heat source of $3.8 \text{ MeV nucleon}^{-1}$ was injected around a depth ${\bar{\rho}=4.3\times10^{11}\text{ g cm}^{-3}}$. However, both studies used different cooling models, which include different assumptions. For example, \citet{turlione2015} fixed the boundary temperature $T_b$ at the base of the envelope during the outburst. Their value of ${3.8 \text{ MeV nucleon}^{-1}}$, is nevertheless within our posterior range (see Figure \ref{fig:histomodel1}).

The most peculiar result that we obtained from modelling the thermal evolution of this source, is that we found indications that the impurity parameter of this neutron star might be higher in part of the crust compared to what has been found for other crust cooling sources. It should be emphasised that the results of model 1 do not provide strong constraints on the impurity parameter throughout the crust. Nevertheless, we found for J1748 that whereas the posterior distribution for the impurity in the outermost region of the crust showed a preference for values on the lower bound of our parameter range (which would be in accordance with other sources), our model suggested a large impurity ($Q_\text{imp}^{(2)}\gtrsim150$, from the grey shaded region in Fig. \ref{fig:histomodel1}) in the second region of the crust ($\rho=10^{11-12}\text{ g cm}^{-3}$, located around the neutron drip). However, the posterior distribution does not allow us to fully exclude the possibility of lower values for $Q_\text{imp}^{(2)}$. 

Finally, our MCMC model 1 showed that core temperature prefers relative high values $\log\tilde T_0\sim8.1$, but low values are not excluded. The envelope composition could not be constrained, and neither could the properties of the inner crust ($Q_\text{imp}^{(3,4,5)}$, superfluid range and dripped neutron entrainment) be probed. The main reason for this is that the base level of the cooling curve is still unknown. Our new observation showed that the observed surface temperature was lower than during previous observation, which indicates that the source is not yet in crust-core equilibrium and thus the base level might not be reached yet. Future observations that constrain the base level and the time at which this level is reached, can constrain the parameters of the crust further, and can provide further insight in the potential high-impurity layer.

In model 1 we do not require the source to cool down to the pre-outburst base level. A change in base level after different outbursts can be explained by variation in envelope composition, determined by the burning processes during accretion. In a second MCMC simulation (model 2) we forced the source to cool down to the measured pre-outburst base level, but left the time at which this level was reached free. This provided constraints on the core temperature, and consequently affected the constraints on other parameters. These results highlight that in order to obtain stronger constraints on the crustal parameters of this source, J1748 needs to be observed in the future such that its base level and the time scale over which this level is reached can be determined.

Although it is clear that we cannot conclusive demonstrate that part of the crust in J1748 needs a high impurity parameter, our results are still suggestive of that and it is interesting to explore what could be the reason for this in this source. Recently, \citet{lau2018} presented nuclear reaction network calculations for accreted neutron star crusts. Assuming various initial compositions (which depend on the thermonuclear burning processes taking place in the envelope), they calculated the reactions that a mass element undergoes as it is pushed deeper into the crust due to continued accretion. Based on the crust composition they made a prediction for the impurity as function of density. They found that the impurity at ${\rho\sim10^9\text{ g cm}^{-3}}$ should have value of a few up to $\sim80$ (depending on the initial composition), and $Q_\text{imp}\sim10$ (for all compositions) just after the neutron drip. As the authors point out, they have included the detailed scattering formalism by \citealt{roggero2016}, which reduces the impurity by a factor 2--4, and consequently the results found using cooling codes that do not include this formalism (as is the case for our code) should be divided by this factor for comparison with the results from \citet{lau2018}. Although our findings do not directly compare with their results, variation in impurity throughout the crust does seem to improve the calculated cooling curves in order  to reproduce the observations. However, a high-impurity region in the crust has not been found before for other sources, although in most cases the impurity is modelled to be constant throughout the crust (except for the deepest layer of the inner crust).

If it is indeed true that J1748 has a relatively low conductivity layer in the crust -- potentially located around the neutron drip -- compared to other sources, it not clear what could cause this difference. \citet{degenaar2013} suggested that the explanation for the difference in cooling behaviour between J1748 and the other sources might be related to several other unusual properties of the source.  First of all, the source is a relatively slow rotating neutron star (with a spin of 11 Hz; \citealt{2010ATel.2929....1S}, see Table 1 in \citealt{degenaar2015ter5x3} for a comparison with other crust cooling sources). Although the neutron star spin has not been measured for all other crust-cooling sources, the sources that do have a measured spin all rotate significantly faster \citep[$>$500 Hz;][]{zhang1998,smith1997,2001ApJ...549L..71W,galloway2010}. However, it is not clear how the spin could affect the heat conductivity and likely it cannot explain the difference in crustal conductivity between  J1748 and the other sources (see the discussion in \citealt{degenaar2013} for more details). 

Secondly, besides the slower spin rate, J1748 might also have a larger surface magnetic field strength \citep[estimated to be $10^{9-10}$ Gauss;][]{papitto2011,papitto2012,miller2011,cavecchi2011} compared to the other sources (no strong constraints exist for these other sources but typically their field strengths are thought to be at most ${10^{8-9}\text{ Gauss}}$). It is unclear how the external magnetic field translates into a crustal field. However, it is possible that the field strength in the crust of J1748 is higher than that in the other sources. In addition, the configuration of the field could also be quite different. Therefore, the  field strength and its configuration in the crust of J1748 could be such that it affects the heat conductivity more strongly than in other cooling sources, causing a lower conductivity layer in J1748.  Our {\sc NSCool} model does not include the possibility of modelling different crustal magnetic field strengths and configurations so we cannot test this hypothesis. 
Moreover, a 2D code is necessary to accurately investigate this because a variety of crustal field strengths and configurations (which are not spherically symmetric) need to be simulated in order to determine if this hypothesis is physically possible (e.g. the required fields might be unrealistically high in order to affect the heat conductivity significantly) and, if so, under what conditions (e.g. maybe it might only be possible when assuming oddly shaped crustal field configuration which might only be present in certain positions in the crust).

Finally, the slow spin and large inferred surface field strength of J1748 compared to the other crust-cooling sources might indicate that the system only relatively recently (within the last 10$^7$ years) started to accrete \citep[e.g.][]{patruno2012}. Therefore, it is possible that the original crust of the neutron star in this system has not completely been replaced by accreted matter. The possibility of the existence of neutron star systems with such hybrid crusts was discussed in detail by \citet[][]{wijnands2013}. \citet[][]{degenaar2013} estimated (using the inferred long-term accretion rate of J1748) that indeed the inner neutron star crust in J1748 might still consist of original matter. Quite possibly the heating processes in and the thermal conductivity of such a crust will differ strongly from that of a purely accreted crust and this could potentially explain why J1748 behaves differently than the other cooling sources. However, calculating the properties of such hybrid crusts are complex as they will change in time because matter keeps being accreted on the neutron star causing more and more of the crust to be replaced. Even if the results of such calculations will be available in the future, it remains difficult to determine exactly how much of the crust in J1748 is replaced and therefore which calculations have to be used to compare our cooling curve to. Recently, \citealt{chaikin2018} presented preliminary results on the thermal evolution of neutron stars with partly accreted crusts and also suggested that this might be applicable to J1748.

In order to determine the exact new base level and to better constrain the impurity parameters in the different crustal layers, additional observations are needed. Since the source is located in a globular cluster, currently only {\it Chandra} has the spatial resolution required to separate the X-ray emission of J1748 from the other cluster sources (see Fig.~\ref{fig:p4:fov}). Unfortunately, the source is already rather faint in our last {\it Chandra} observations and it requires long exposure times to get reasonable X-ray spectra. The July 2016 observations have a total exposure time of $\sim$130 ksec and similar exposure times are needed in future {\it Chandra} observations to accurately constrain the temperature (and likely longer since the source is expected to cool further resulting in a lower count rate). Furthermore, additional observations are only really constraining in $>$2000 d from now, so in about 5 yrs. Observations performed earlier will improve the statistics of our current modelling constraints on the crustal parameters, but will not conclusively determine whether the source will cool down to a new base level or to the temperature measured pre-outburst.

\section*{Acknowledgements}

The authors would like to thank the referee for detailed and constructive comments that helped to improve the paper. S.V. would like to acknowledge support from NOVA (Nederlandse Onderzoekschool Voor Astronomie). L.O., R. W., and A.P. acknowledge support from a NWO Top Grant, module 1 (awarded to R.W.). N.D. is supported by a Vidi grant from NWO. D.P's work is partially supported by Conacyt through the Fondo Sectorial de Investigaci\'{o}n para la Educaci\'{o}n, grant CB-2014-1, No. 240512. J.H. and J.M.M. acknowledge financial support through Chandra Award Number GO6-17031B issued by the Chandra X-ray Observatory Center, which is operated by the Smithsonian Astrophysical Observatory for and on behalf of the National Aeronautics Space Administration under contract NAS8-03060. This research has made use of the MAXI data provided by RIKEN, JAXA and the MAXI team.





\bibliographystyle{mnras}
\bibliography{ter5_x2} 



\appendix
\begin{onecolumn}


 
\section{Results of observational analysis assuming different masses and radii} 
\label{section:appMRanalysis}

\begin{table*}
\setlength{\tabcolsep}{3pt}
\caption{Quiescent temperatures obtained from spectral analysis for nine different assumed mass-radius combinations. The assumed neutral hydrogen column density for all spectral  fits is $N_\text{H}=3.07\pm 0.07\times10^{22}\text{ cm}^{-2}$.}
 \label{tab:app3-temps}
 \begin{tabular}{cccccccccc}
 \hline
 \multicolumn{10}{c}{$kT_\text{eff}^{\infty}$ (eV)} \tabularnewline
 \tabularnewline
 & & $M=1.0 \text{ M}_\odot$ & & & $M=1.4 \text{ M}_\odot$ & & & $M=1.8 \text{ M}_\odot$ & \tabularnewline
  MJD & R = 9 km & R = 12 km & R = 15 km & R = 9 km & R = 12 km & R = 15 km & R = 9 km & R = 12 km & R = 15 km\tabularnewline \cmidrule(l{3pt}r{3pt}){1-1} 
  \cmidrule(l{3pt}r{3pt}){2-4} \cmidrule(l{3pt}r{3pt}){5-7} \cmidrule(l{3pt}r{3pt}){8-10}
 52833.6/55027	&  82.7$\pm$2.0 &  77.6$\pm$1.9 &  74.1$\pm$1.7 & 	 80.1$\pm$1.9 	& 	 75.9$\pm$1.8 	& 	 72.5$\pm$1.6 	 & 	 77.2$\pm$1.8 & 	 74.2$\pm$1.8 	& 	 71.1$\pm$1.7 	  \tabularnewline
55609.4 	& 115.3$\pm$1.7&  107.9$\pm$1.5&  102.2$\pm$1.4 & 	111.4$\pm$1.6 	& 	105.0$\pm$1.4 	& 	100.1$\pm$1.4 	 & 	107.0$\pm$1.5& 	 	102.2$\pm$1.4 & 	 98.0$\pm$1.3 	  \tabularnewline
55680.7 	& 104.4$\pm$1.7&   98.0$\pm$1.6 &  93.0$\pm$1.5 & 	100.7$\pm$1.6 	& 	 95.5$\pm$1.5 	& 	 91.1$\pm$1.4 	 & 	 97.0$\pm$1.6 & 	 93.1$\pm$1.4 	& 	 89.4$\pm$1.4 	  \tabularnewline
55812.7 	& 102.8$\pm$1.6&   96.4$\pm$1.5 &  91.4$\pm$1.4 & 	 99.1$\pm$1.6 	& 	 94.0$\pm$1.5 	& 	 89.6$\pm$1.4 	 & 	 95.4$\pm$1.5 & 	 91.6$\pm$1.4 	& 	 87.8$\pm$1.4 	  \tabularnewline
56060.8 	&  96.3$\pm$1.9 &  90.5$\pm$1.7 &  86.1$\pm$1.6 & 	 93.0$\pm$1.8 	& 	 88.4$\pm$1.7 	& 	 84.4$\pm$1.6 	 & 	 89.5$\pm$1.8 & 	 86.3$\pm$1.6 	& 	 82.9$\pm$1.5 	  \tabularnewline
56207.9 	&  98.5$\pm$1.6 &  92.5$\pm$1.5 &  87.8$\pm$1.4 & 	 95.0$\pm$1.5 	& 	 90.2$\pm$1.4 	& 	 86.1$\pm$1.4 	 & 	 91.5$\pm$1.5 & 	 88.0$\pm$1.4 	& 	 84.5$\pm$1.3 	  \tabularnewline
56336.8 	&  95.3$\pm$0.9 &  89.5$\pm$0.9 &  84.9$\pm$0.8 & 	 92.1$\pm$0.9 	& 	 87.3$\pm$0.9 	& 	 83.2$\pm$0.8 	 & 	 88.5$\pm$0.9 & 	 85.1$\pm$0.8 	& 	 81.6$\pm$0.8 	  \tabularnewline
56489.9 	&  96.9$\pm$2.4 &  91.0$\pm$2.2 &  86.6$\pm$2.1 & 	 93.6$\pm$2.2 	& 	 88.9$\pm$2.1 	& 	 84.9$\pm$2.0 	 & 	 90.0$\pm$2.2 & 	 86.7$\pm$2.0 	& 	 83.3$\pm$2.0 	  \tabularnewline
56855.8 	&  94.2$\pm$1.4 &  88.0$\pm$1.3 &  83.3$\pm$1.2 & 	 90.8$\pm$1.3 	& 	 85.9$\pm$1.2 	& 	 81.7$\pm$1.2 	 & 	 87.0$\pm$1.2 & 	 83.7$\pm$1.2 	& 	 80.1$\pm$1.1 	  \tabularnewline
57583.6 	&  90.3$\pm$1.3 &  84.5$\pm$1.2 &  80.0$\pm$1.2 & 	 87.2$\pm$1.3 	& 	 82.4$\pm$1.2 	& 	 78.5$\pm$1.1 	 & 	 83.7$\pm$1.2 & 	 80.4$\pm$1.1 	& 	 77.0$\pm$1.1 	  \tabularnewline  \hline
 \end{tabular}
\end{table*}

\begin{table*}
\centering
\caption{Overview of the surface gravity and crust radius for the nine different mass-radius combinations for which the observational data was reanalysed.}
\label{tab:gs14deltaR}
\begin{tabular}{ccccccc}
\hline
                 & \multicolumn{2}{c}{$R=9$ km}                          & \multicolumn{2}{c}{$R=12$ km}                         & \multicolumn{2}{c}{$R=15$ km}                         \\
M ($\text{M}_\odot$) & $g_\text{s}\; (10^{14}\text{ cm s}^{-2})$ & $\Delta R$ (km) & $g_\text{s}\; (10^{14}\text{ cm s}^{-2})$ & $\Delta R$ (km) & $g_\text{s}\; (10^{14} \text{ cm s}^{-2})$ & $\Delta R$ (km)  \\ 
\cmidrule(l{3pt}r{3pt}){1-1} \cmidrule(l{3pt}r{3pt}){2-3} \cmidrule(l{3pt}r{3pt}){4-5} \cmidrule(l{3pt}r{3pt}){6-7}
1.0              & 2.00                                & 0.96            & 1.06                                & 1.83            & 0.66                                & 2.93            \\
1.4              & 3.12                                & 0.58            & 1.59                                & 1.20            & 0.97                                & 2.01            \\
1.8              & 4.61                                & 0.35            & 2.22                                & 0.81            & 1.32                                & 1.43            \\ \hline
\end{tabular}
\end{table*}
 \begin{figure}
 \begin{center}
 \includegraphics[width=0.6\textwidth]{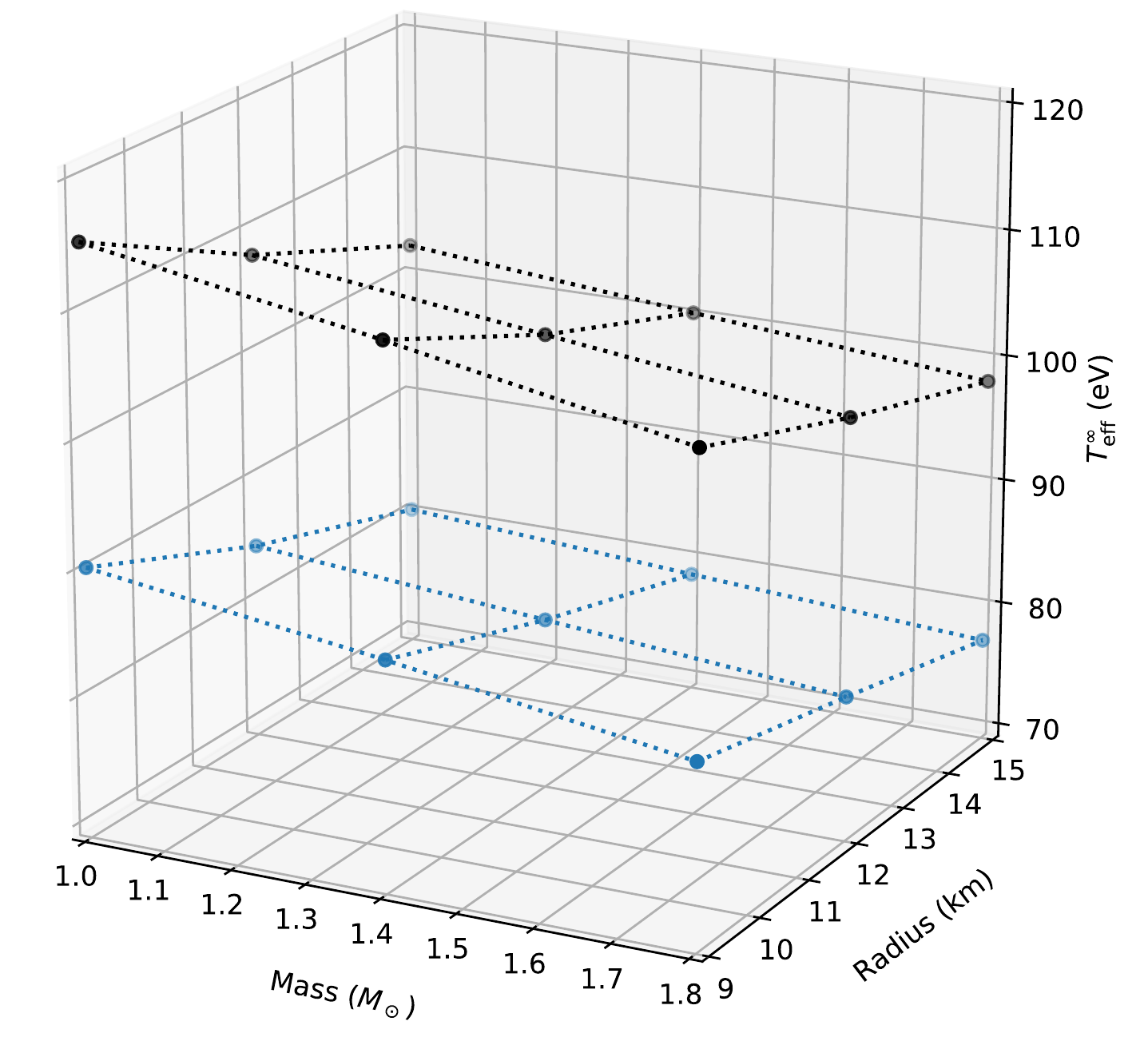}
 \caption{3-dimensional representation of the temperatures of the first (black points) and last (blue points) observation for different assumed masses and radii. The surfaces formed by the temperatures of intermediate observations are omitted for clarity. }
 \label{fig:MR-temp-plane}
 \end{center}
 \end{figure}
 
In this section, we present the results of the reanalysis of the observations for different assumed neutron star masses and radii. We reanalysed the spectral data assuming nine different mass-radius combinations. We assumed masses of 1.0, 1.4, and ${1.8\text{ M}_\odot}$, corresponding to light, intermediate and heavy neutron stars, and radii of 10, 12, and 15 km. The resulting mass-radius combinations lead to stars that vary strongly in compactness and crust radius (see Table. \ref{tab:gs14deltaR}).

We reanalysed the observations in the same way as described in Section \ref{sec:p4:obs}. The only difference is that we assumed for all spectral fits presented in this section a hydrogen column density ${N_\text{H}=(3.07\pm 0.07)\times10^{22}\text{ cm}^{-2}}$. For each mass-radius combination the $N_\text{H}$ was determined from an \textsc{nsatmos} fit using all spectra. The assumed $N_\text{H}$, is the average of these values. Assuming a constant $N_\text{H}$ throughout the spectral fits assures that the variation in resulting surface temperatures is only caused by a difference in assumed mass and radius. The obtained temperatures are presented in Table \ref{tab:app3-temps}. In the MCMC simulations, we use linear interpolation from this grid (see Fig. \ref{fig:MR-temp-plane} for the surfaces, which are almost planar, formed by the temperatures of one observation as function of mass and radius) to obtain the temperatures for any specific mass-radius combination.

\clearpage
\begin{figure}
\section{Additional figures of MCMC simulations}\label{section:appMCMC}
\begin{center}
\includegraphics[width=0.65\textwidth]{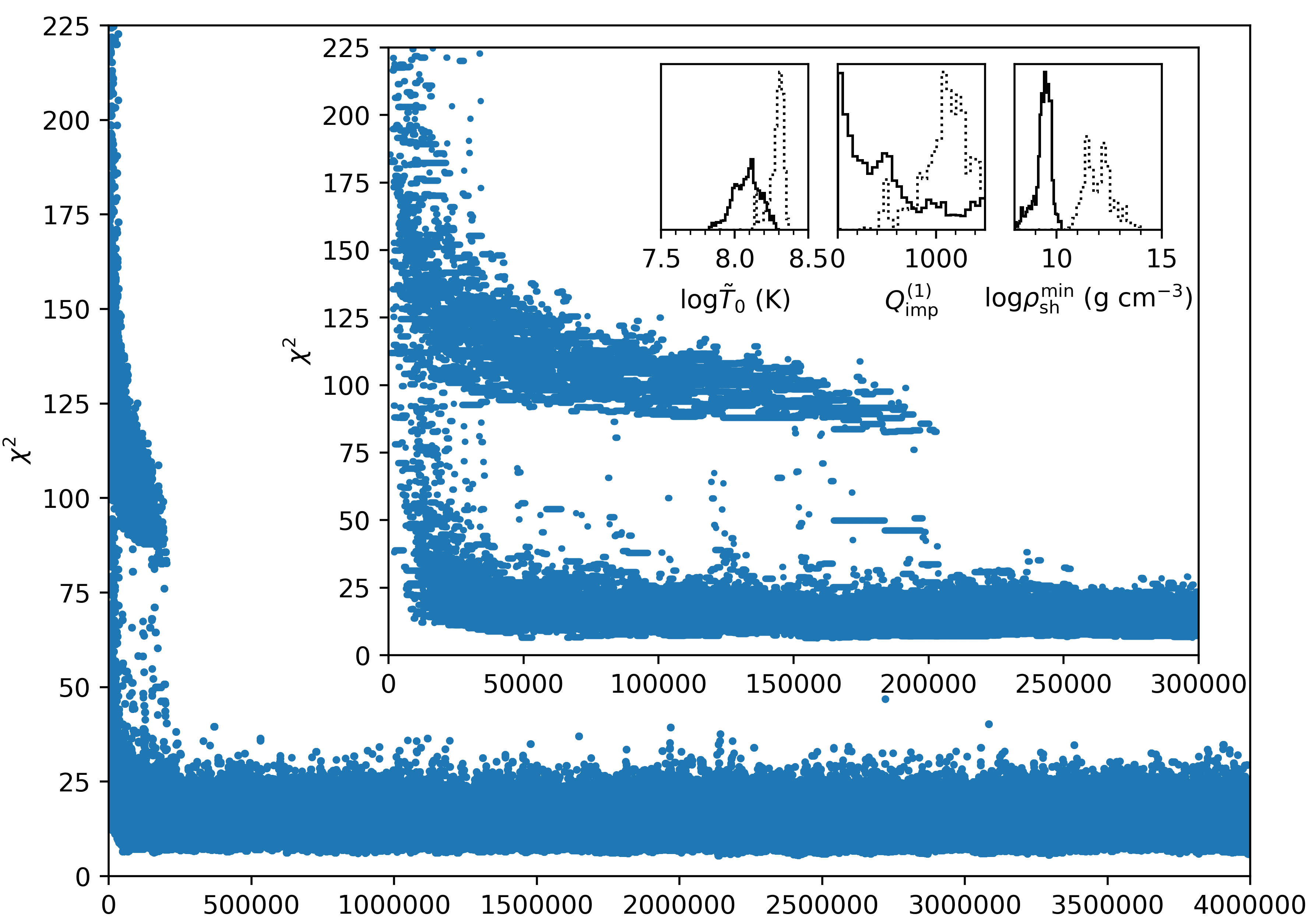}
\caption{ Scatter plot of $\chi^2$ in our model 2 run showing a short initial burn-in phase of $\sim$30'000 points during which all $\chi^2$ drop below 200, but exhibiting part of the walkers trapped in a local minimum with $\chi^2 > 75$ till more than 200'000 points have been generated. The inset shows the first 300'000 points in more detail and within it histograms of $T_0$, $Q_\mathrm{imp}^{(1)}$ and $\log \rho_\mathrm{sh, \, min}$ from the points 50'000 to 200'000 with $\chi^2 < 75$ as continuous lines and with $\chi^2 > 75$ as doted lines clearly exhibiting the bi-modality. This {\em seems} to indicate that after about 300'000 points the chain has converged, but the next figure shows that this likely required 1'000'000 points. } 
\label{fig:AppMCchi2bimodal}
\end{center}
\end{figure}

\begin{figure}
\begin{center}
\includegraphics[width=0.43\textwidth]{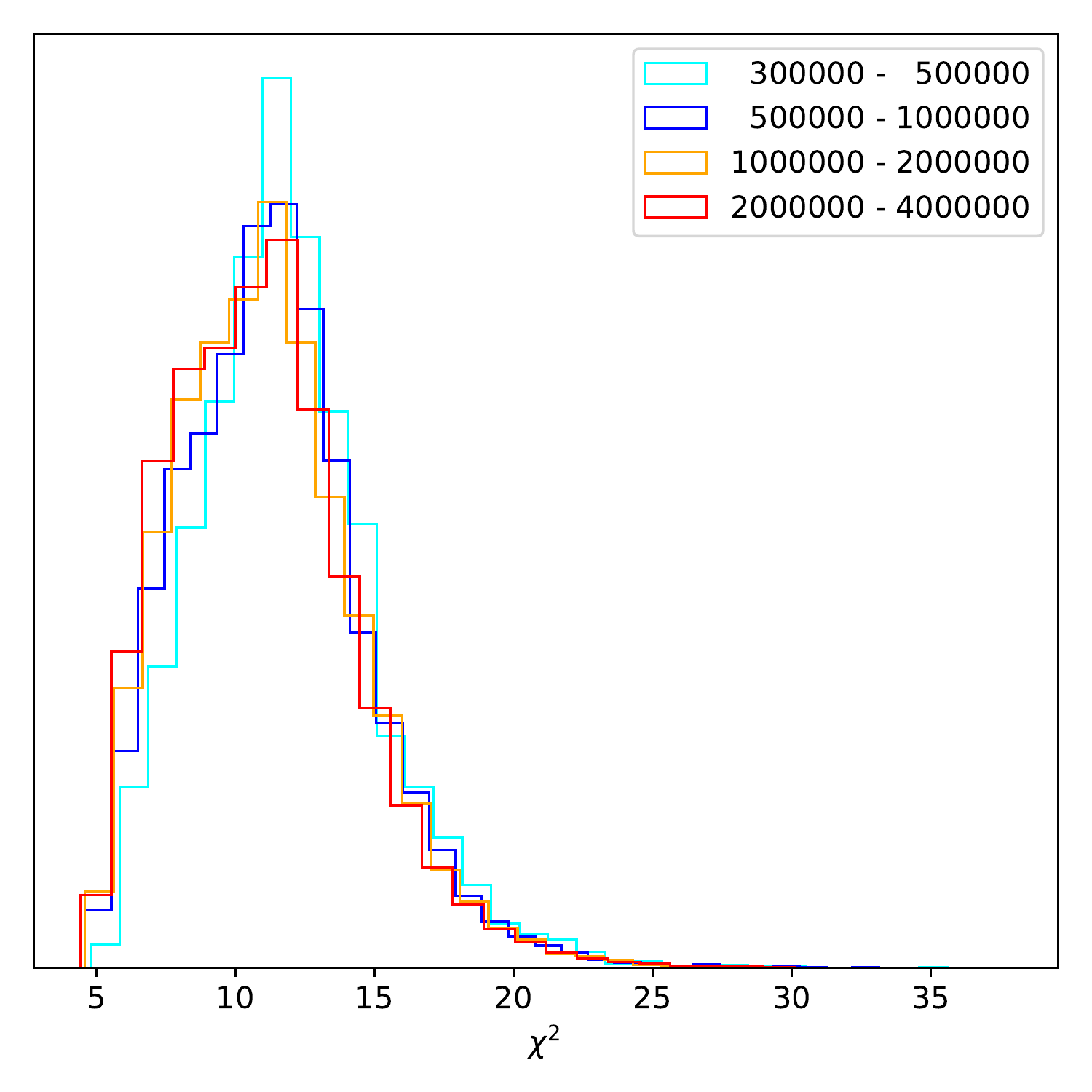}
\includegraphics[width=0.43\textwidth]{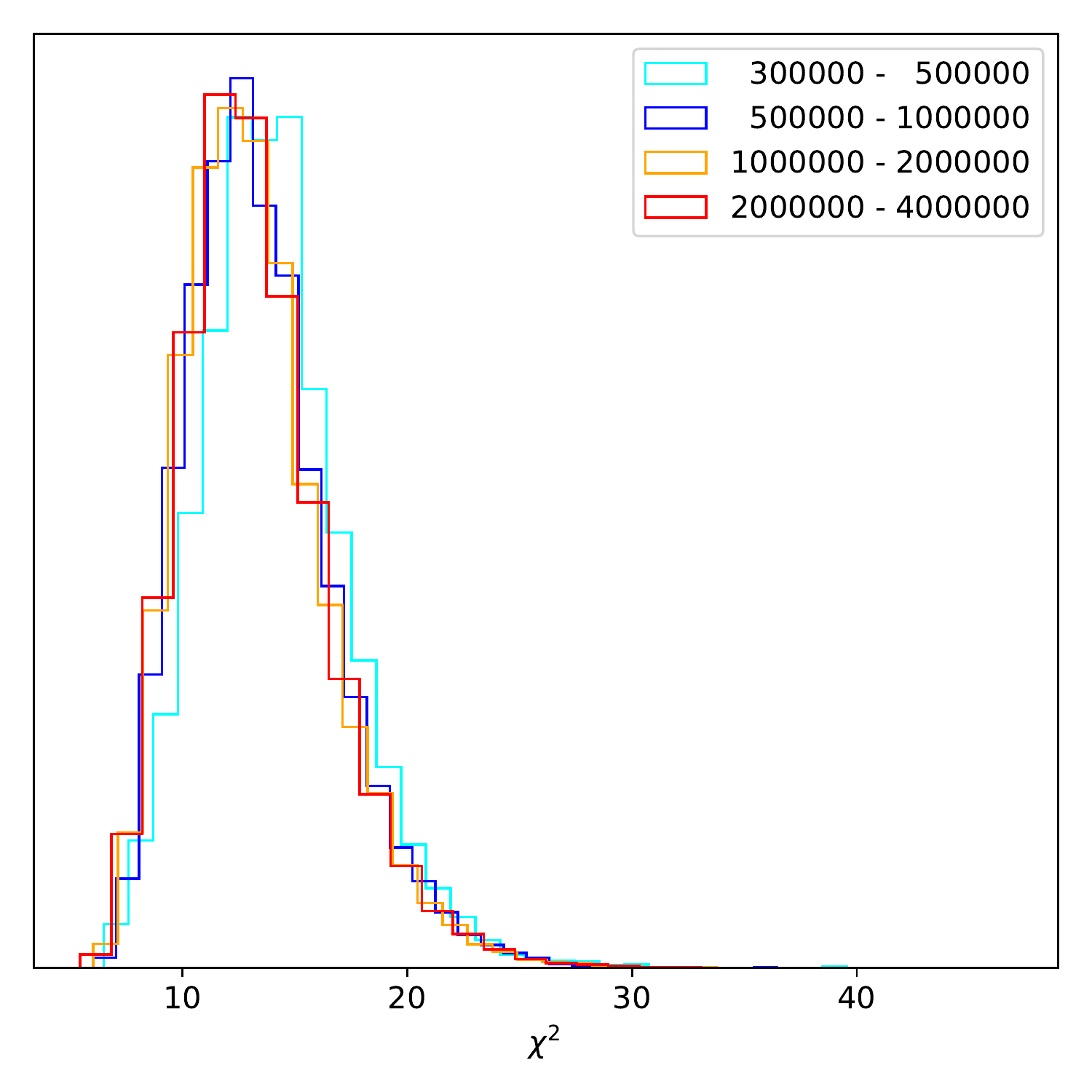}
\caption{The $\chi^2$ distributions for selected number points in our MCMC run for model 1 (left) and model 2 (right). As there seems to be no significant change in the distributions after 1 million points (for both panels), we concluded that convergence seemed to be reached at that stage. We therefore eliminated the initial 1 million curves as "burn-in phase" from our total of 4 million cooling curves.} 
\label{fig:AppMCchi2}
\end{center}
\end{figure}

\end{onecolumn}

\bsp	
\label{lastpage}

\end{document}